\documentclass[a4paper,fleqn]{cas-dc}

\usepackage[numbers,sort&compress]{natbib}

\usepackage{graphicx}%
\usepackage{multirow}%
\usepackage{amsmath,amssymb,amsfonts}%
\usepackage{amsthm}%
\usepackage{mathrsfs}%
\usepackage[title]{appendix}%
\usepackage{xcolor}%
\usepackage{textcomp}%
\usepackage{manyfoot}%
\usepackage{booktabs}%
\usepackage{algorithm}%
\usepackage{algorithmicx}%
\usepackage{algpseudocode}%
\usepackage{listings}%
\usepackage{float} 
\usepackage{subfigure}
\usepackage{threeparttable}
\usepackage[figuresright]{rotating}
\usepackage{caption}

\usepackage[defaultcolor=cyan]{changes}

\def\tsc#1{\csdef{#1}{\textsc{\lowercase{#1}}\xspace}}
\tsc{WGM}
\tsc{QE}

\newdefinition{rmk}{Definition}

\begin{document}
\let\WriteBookmarks\relax
\def\floatpagepagefraction{1}
\def\textpagefraction{.001}

\shorttitle{Advances and Challenges of Multi-task Learning Method in Recommender Systems: A Survey}    

\shortauthors{Zhang et al.}

\title[mode = title]{Advances and Challenges of Multi-task Learning Method in Recommender Systems: A Survey}

\author[1]{Mingzhu Zhang}
\ead{zhangmz@emails.bjut.edu.cn}
\credit{Conceptualization of this study, Methodology, Software}

\author[1]{Ruiping Yin}
\cormark[1] 
\ead{yinruiping@bjut.edu.cn}
\credit{Conceptualization of this study, Methodology, Software}

\author[1]{Zhen Yang}
\ead{yangzhen@bjut.edu.cn}
\credit{Conceptualization of this study, Methodology, Software}

\author[1]{Yipeng Wang}
\ead{wangyipeng@bjut.edu.cn}
\credit{Conceptualization of this study, Methodology, Software}

\address[1]{Beijing University of Technology, Beijing, 100124, China}

\cortext[1]{Corresponding author} 

\begin{abstract}
Multi-task learning (MTL) has been widely applied to modern RSs, enabling the simultaneous optimization of diverse user feedback such as clicks, purchases, and dwell time. Despite its success, MTRSs face significant challenges, most notably negative transfer and data sparsity in long-tail behaviors. To provide a holistic view of solutions to these challenges, this survey presents a systematic review of the MTRSs landscape. We first articulate the motivation behind MTRSs, analyzing how task dependencies and conflicts necessitate specialized modeling. Subsequently, we introduce a hierarchical taxonomy centered on two pillars: (1) Architecture Paradigms, which classify methods into parameter sharing (e.g., Mixture-of-Experts) to alleviate interference, transfer learning (e.g., cascading structures) to address sparsity, adversarial learning for robust representation, and the recent LLM-based architectures; and (2) Optimization Strategies, which examine loss balancing and gradient conflict techniques to achieve Pareto optimality.(3) Training Paradigms, which explore pre-training, reinforcement learning and auxiliary objectives to improve data efficiency in MTRSs. Finally, we provide an outlook on open problems and future trends.
\end{abstract}




\maketitle

\section{Introduction}

With the explosive growth of data in the information age, Recommender systems (RSs) have emerged as an indispensable tool to help users filter useful or interesting content from massive information streams. They participate in almost every aspect of modern life, such as E-commerce recommendation (e.g., Taobao and Amazon), multimedia services (e.g., Netflix, Spotify and TikTok), and social services (e.g., WeChat and Facebook). A highly effective recommender system that is both accurate and timely can provide significant value to both users and businesses. Consequently, the development of recommendation techniques continues to attract substantial academic and industrial attention, with the aim of improving the quality of recommendations, enhancing user satisfaction, and supporting a wide range of business objectives.

Traditional recommendation methods, especially collaborative filtering (CF)~\cite{goldberg1992using} and content-based filtering, have been the mainstream of RSs for many years. CF relies on user-item interactions to recommend items based on historical behavior patterns of similar users or items, while content-based methods recommend items by analyzing item metadata and user profiles. In addition, hybrid RSs that combine CF and content-based signals have been widely explored to leverage the strengths of both paradigms. Despite their success, these traditional methods still suffer from several inherent limitations. First, the \emph{cold-start} problem remains challenging: for new users or items with little interaction history, CF-based models struggle to make accurate predictions, and although content-based or hybrid approaches can exploit side information, they often fail to capture fine-grained, long-term preference patterns. Second, real-world user-item interaction matrices are typically highly sparse, with most users only interacting with a small fraction of available items, which severely restricts the ability to learn reliable collaborative signals and leads to suboptimal recommendations. Moreover, logged interaction data are generated under historical exposure and recommendation policies, which introduces \emph{selection bias} and makes it difficult for traditional models to accurately recover users' true preferences over the entire candidate space.

In recent years, data-driven deep learning technologies have significantly advanced data mining and information retrieval, particularly in RSs. Deep models and pre-trained embedding techniques such as Word2Vec have markedly improved user preference prediction and recommendation accuracy, as demonstrated by models such as DeepFM~\cite{guo_deepfm_2017}, DSTNs~\cite{ouyang_deep_2019}, and DIEN~\cite{zhou_deep_2019}. However, most of these approaches are still trained in a single-task manner, relying on one primary supervision signal (e.g., clicks). They require large volumes of interaction data and user feedback, which are often difficult to obtain, especially for fine-grained or long-term objectives (e.g., conversion or retention). Moreover, deploying separate deep models for different objectives across diverse datasets demands substantial computational resources, time, and maintenance overhead. These limitations motivate the community to explore integrating Multi-Task Learning (MTL) into RSs as a principled way to jointly exploit multiple behavioral signals.

\begin{figure}
    \centering
    \includegraphics[width=1\linewidth]{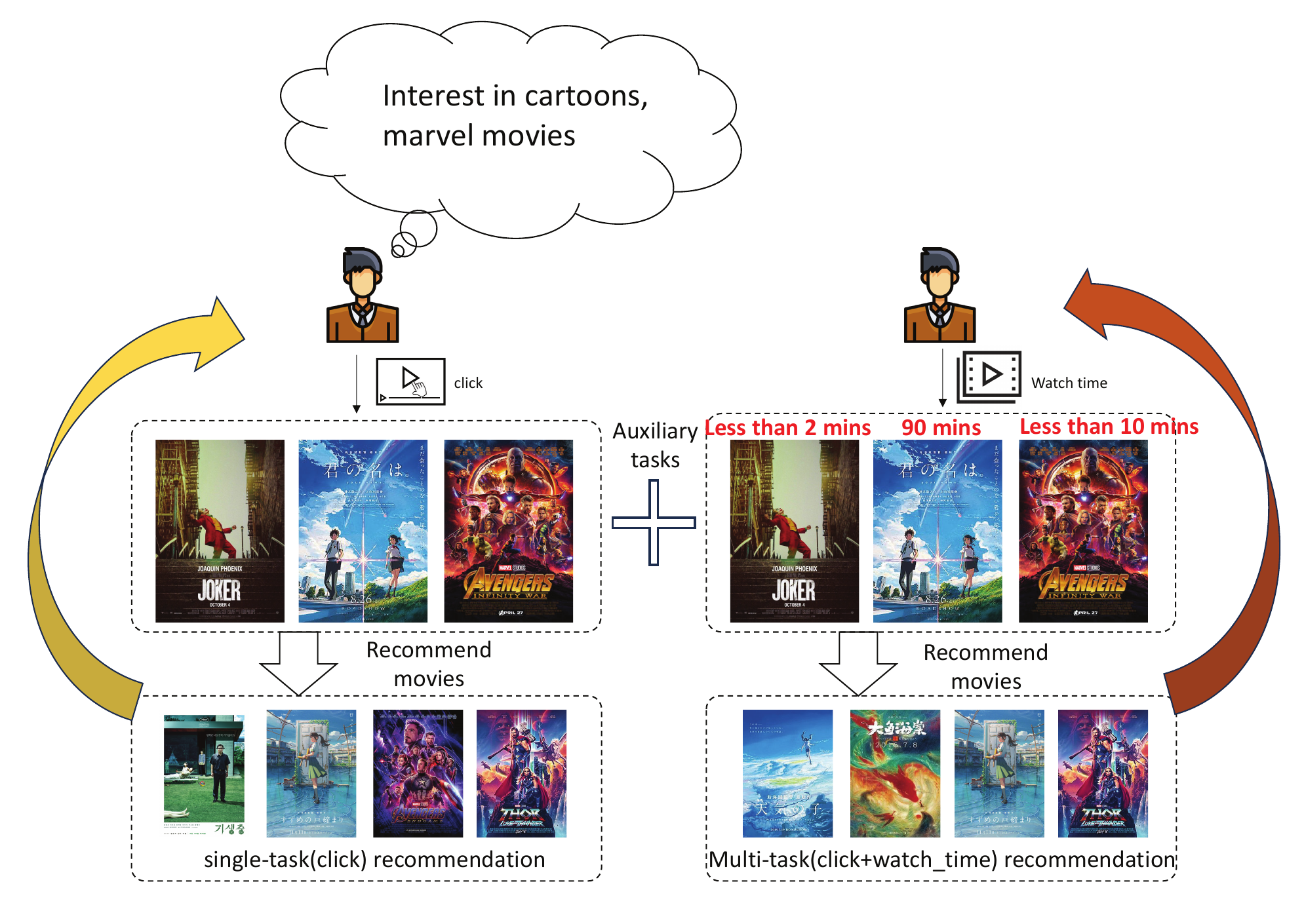}
    \caption{A illustration example of traditional single-task recommendation vs multi-task recommendation}
    \label{intro}
\end{figure}

The advent of MTL brings new energy to RSs by promoting information sharing and the simultaneous optimization of multiple objectives. As illustrated in Fig.~\ref{intro}, if recommendations are solely based on user click behavior, there is a high risk of recommending movies similar to ''Joker'' merely because the user clicked on it. However, this interaction may not necessarily reflect the user's true preference; the user might have clicked on ''Joker'' due to its high rating or extensive promotion rather than an inherent interest in that genre. By incorporating additional tasks, such as modeling viewing duration, dwell time, or long-term satisfaction, MTL enables a more holistic and robust understanding of user interests compared to traditional single-task learning. This intuitive example suggests that leveraging multi-task recommendation approaches, which jointly consider various user-related signals, can substantially improve recommendation precision.

From a learning-theoretic perspective, the benefits of MTL in RSs are supported by classical analyses showing that jointly learning related tasks introduces an inductive bias that constrains the hypothesis space and reduces generalization error \cite{mitchell_multitask_1997, baxter2000model, ruder_overview_2017}. Auxiliary objectives act as implicit regularizers, encouraging the model to discover shared structure across tasks rather than overfitting to noisy single-task supervision. This makes MTL particularly attractive for recommendation scenarios, where signals are often sparse, biased, and heterogeneous. A more detailed discussion of these theoretical motivations is provided in Section~\ref{sec2}.

In addition to such intuitive examples, a growing body of empirical work quantitatively demonstrates the effectiveness of MTL in large-scale RSs. For instance, ESMM~\cite{ma_entire_2018} shows that jointly modeling Click-Through Rate (CTR) and Conversion Rate (CVR) tasks over the entire impression space can effectively alleviate sample selection bias, achieving a 2.18\% AUC gain on CVR prediction. PLE~\cite{tang_progressive_2020} reports consistent improvements on conflicting objectives (e.g., view-through rate vs. view completion rate) by disentangling shared and task-specific components. In industrial deployments, large-scale platforms such as Google \cite{ma_modeling_2018}, Kuaishou \cite{wang2025home}, and YouTube \cite{zhao_recommending_2019} have adopted multi-task recommendation models and consistently observed significant lifts in CVR and gross merchandise volume (GMV), thereby empirically validating the practical value of MTL in real-world recommender and advertising systems.

Therefore, incorporating MTL techniques helps address several core challenges faced by RSs (see Section~\ref{sec2} for a more detailed analysis):

\begin{itemize}
    \item \textbf{Alleviating data sparsity and cold start}: Dense objectives such as CTR or watch-time provide rich supervision to enhance sparse targets like CVR, add-to-cart, or long-term retention, thereby improving performance in data-scarce regimes.
    \item \textbf{Mitigating selection bias}: By jointly modeling multiple stages in the user decision process (e.g., impression~$\rightarrow$ click~$\rightarrow$ conversion), MTL can partially correct for sample selection bias induced by historical logging policies.
    \item \textbf{Improving efficiency and maintainability}: Sharing embeddings and lower layers across tasks reduces parameter redundancy and consolidates multiple single-task pipelines into a unified architecture, lowering both computation and engineering maintenance costs.
    \item \textbf{Enhancing generalization}: Auxiliary objectives act as regularizers that encourage the model to learn more robust and transferable representations, leading to better generalization on the primary tasks.
\end{itemize}

\textbf{What distinguishes this survey from earlier ones? } As summarized in Table~\ref{tab0}, numerous surveys have been proposed on MTL in general~\citep{zhang_overview_2018,thung_brief_2018,zhang_survey_2022,vandenhende_multi-task_2022,liebel_auxiliary_2018,chen_multi_2021,worsham_multi_2020} and on RSs from different perspectives~\citep{zhang_deep_2020,lin_survey_2022,gao_advances_2021,jannach_survey_2022,deldjoo_survey_2022,zhu_cross-domain_2021,liu_multimodal_2023,chen_deep_2023,gao_survey_2023,wu_graph_2023,zheng_automl_2023}. Most of these works either focus on generic MTL methodology with limited coverage of recommendation applications, or concentrate on specific RS paradigms (e.g., deep learning, reinforcement learning, cross-domain RS, or conversational RS) while only briefly touching on multi-objective or multi-task modeling.

In addition, several recent surveys specifically focus on \emph{multi-behavior recommendation}~\citep{chen2023survey,kim2025multi},summarizing architectures that exploit heterogeneous user feedback and behavior interaction patterns. However, these works typically treat MTL only implicitly and do not provide a systematic taxonomy of MTL-specific issues such as task relations, gradient conflicts, or the coupling between parameter-sharing architectures and optimization strategies.

Recently, Wang et al. \cite{wang2023multi} provide a survey on \emph{multi-task deep RSs}, mainly categorizing existing models by architectural design. While highly valuable, their work pays less attention to (i) how MTL relates to broader recommendation challenges such as multi-scenario and multi-domain settings, and (ii) the rapidly emerging line of MTL with large pre-trained foundation models and generative recommendeation. In contrast, this survey aims to provide a more unified and forward-looking treatment of \emph{multi-task learning-based RSs} (MTRSs) by jointly analyzing task relations, model architectures (including LLM-based MTRSs), optimization strategies, and learning paradigms, and by explicitly linking them to core recommendation challenges.

\begin{figure}
\centering  
\subfigure[Distribution of MTRSs Publications in Recent 7 Years]{
\label{dis.sub.1}
\includegraphics[width=0.45\textwidth]{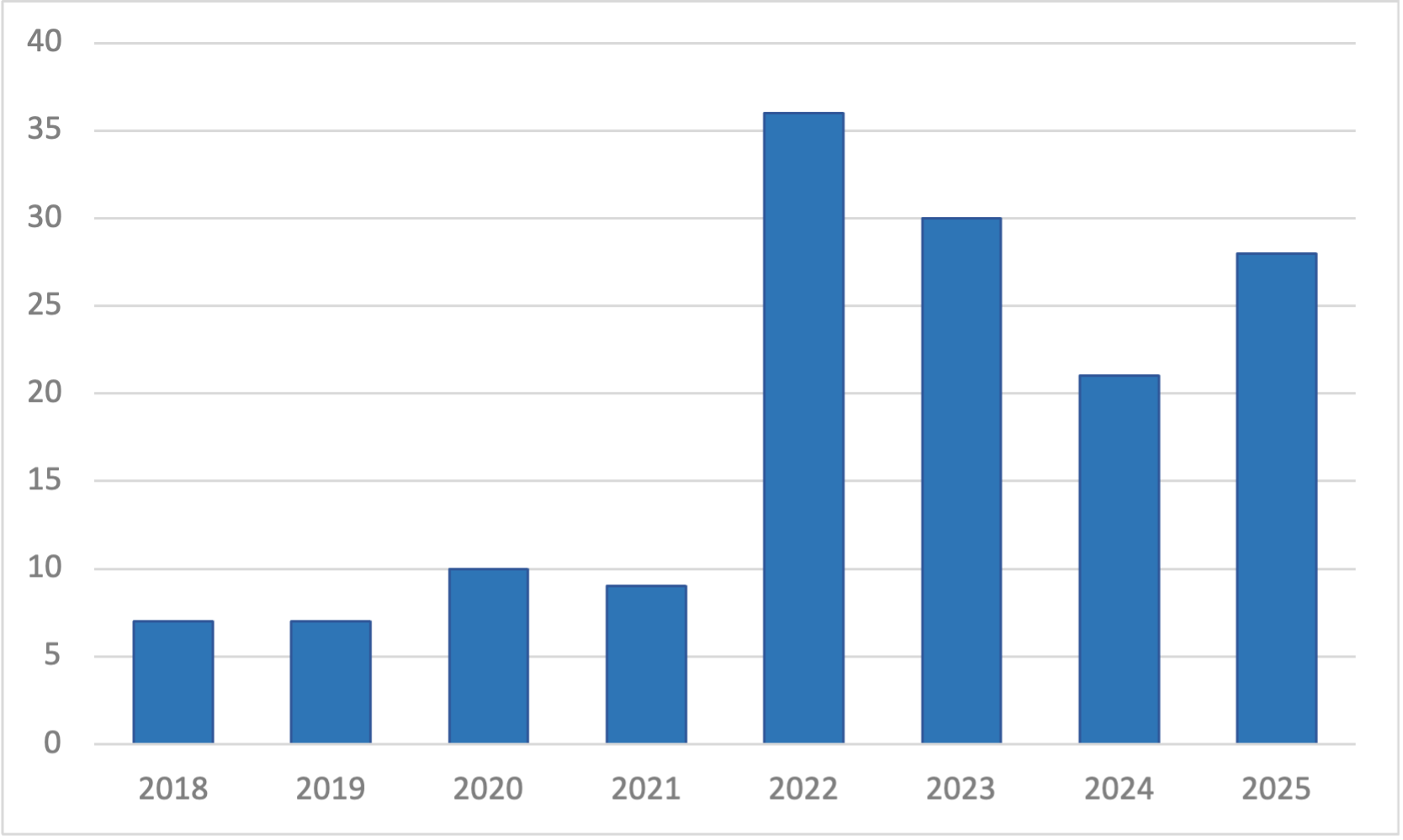}}
\subfigure[Venue Distribution of Published MTRSs in Recent 3 Years.]{
\label{dis.sub.2}
\includegraphics[width=0.45\textwidth]{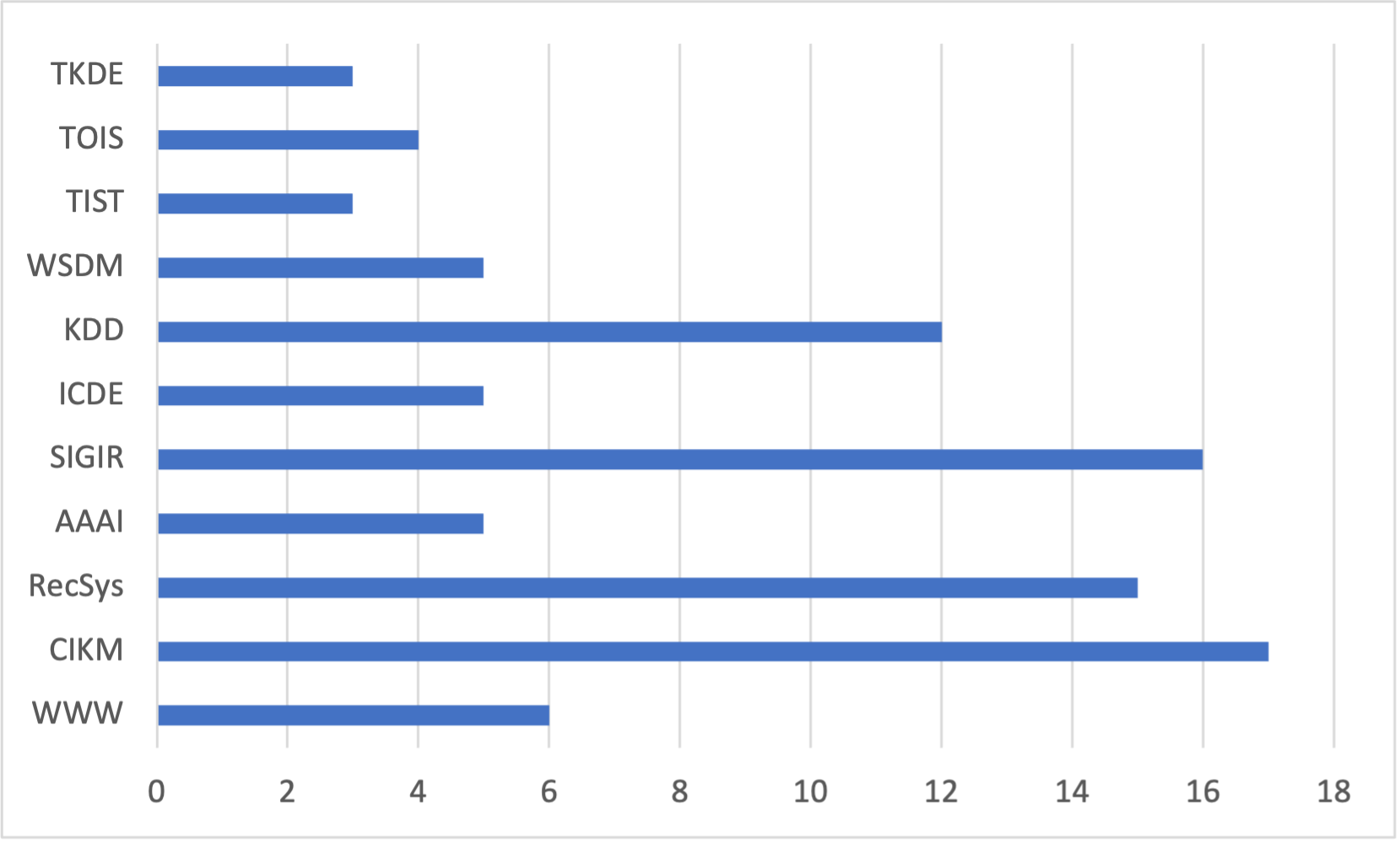}}
\caption{The Statistical Analysis of Publications of MTRSs in Recent Years}
\label{dis}
\end{figure}

This survey seeks to adopt a broader perspective to unify the analysis of MTRSs and their associated problems, systematically distill the fundamental modeling principles, and, in light of recent research progress and practical challenges, propose a more generalized and forward-looking research framework and outlook. We reviewed and selected publications from related high-profile conferences and journals (e.g., AAAI, CIKM, ICDE, IJCAI, RecSys, KDD, SIGIR, TKDE, WSDM, WWW, etc.) to discover more recent works, including representative papers published up to 2025. As a result, over 100 studies were shortlisted and classified in this survey. Furthermore, we conducted a statistical analysis of MTRS publications and observed that the application of MTL in RSs is not limited to industry but has also become an active and growing trend in academia, as illustrated in Fig.~\ref{dis}.

\begin{table*}[h]
\centering
\caption{Related surveys of MTL and RSs.}
\label{tab0}
\resizebox{\textwidth}{!}{
\begin{tabular}{ccccccccccccccc}
\hline
\multirow{2}{*}{Categories} & \multicolumn{1}{c}{\multirow{2}{*}{Surveys}}       & \multicolumn{4}{c}{RSs type}                                                                                                             & \multicolumn{4}{c}{MTL techniques}                                                                            & \multicolumn{3}{c}{DL techniques}                                                 & \multirow{2}{*}{Datasets} & \multirow{2}{*}{Evaluation metric} \\ \cmidrule(r){3-6} \cmidrule(r){7-10}\cmidrule(r){11-13}& \multicolumn{1}{c}{}                               & \multicolumn{1}{c}{CDR} & \multicolumn{1}{c}{IRSs} & \multicolumn{1}{c}{CRs} & \multicolumn{1}{c}{ERSs} & Task relation             & Architecture              & Optimization              & Application               & GNN                       & RL                        & KG                        &                           &                                    \\ \cline{1-13}
\multirow{9}{*}{RSs}        & \cite{lin_survey_2022}         &                                  & \checkmark       & \checkmark        & \checkmark        &                           &                           &                           &                           &                           & \checkmark &                           &                           &                                    \\
                            & \cite{afsar_reinforcement_2022} &                                  & \checkmark       &                                  &                                  &                           &                           &                           &                           &                           & \checkmark & \checkmark &                           &                                    \\
                            & \cite{gao_advances_2021}        &                                  &                                 & \checkmark        &                                  &                           &                           &                           &                           &                           &                           &                           & \checkmark & \checkmark          \\
                            & \cite{jannach_survey_2022}      &                                  &                                 & \checkmark        &                                  &                           &                           &                           &                           &                           &                           &                           & \checkmark & \checkmark          \\
                            & \cite{zhu_cross-domain_2021}    & \checkmark        &                                 &                                  &                                  &                           &                           &                           &                           &                           &                           &                           &                           &                                    \\
                            & \cite{liu_multimodal_2023}      &                                  & \checkmark       &                                  &                                  &                           &                           &                           &                           &                           &                           & \checkmark & \checkmark &                                    \\
                            & \cite{chen_deep_2023}           &                                  & \checkmark       &                                  &                                  &                           &                           &                           &                           &                           & \checkmark &                           &                           &                                    \\
                            & \cite{gao_survey_2023}          & \checkmark        & \checkmark       & \checkmark        & \checkmark        &                           &                           &                           &                           & \checkmark &                           & \checkmark &                           &                                    \\
                            & \cite{wu_graph_2023}            &                                  & \checkmark       &                                  & \checkmark        &                           &                           &                           &                           & \checkmark &                           & \checkmark & \checkmark & \checkmark          \\ \hline
\multirow{7}{*}{MTL}        & \citep{zhang_overview_2018}     &                                  &                                 &                                  &                                  & \checkmark & \checkmark &                           & \checkmark &                           & \checkmark &                           &                           &                                    \\
                            & \cite{ruder_overview_2017}      &                                  &                                 &                                  &                                  & \checkmark & \checkmark &                           &                           &                           &                           &                           &                           &                                    \\
                            & \cite{thung_brief_2018}         &                                  &                                 &                                  &                                  &                           & \checkmark & \checkmark & \checkmark &                           &                           &                           &                           &                                    \\
                            & \cite{zhang_survey_2022}        &                                  &                                 &                                  &                                  & \checkmark & \checkmark & \checkmark &                           &                           &                           &                           & \checkmark &                                    \\
                            & \cite{vandenhende_multi-task_2022}             &                                 &                                  &                                  &                           & \checkmark & \checkmark &                           &                           &                           &                           & \checkmark & \checkmark          \\
                            & \cite{chen_multi_2021}          &                                  &                                 &                                  &                                  & \checkmark & \checkmark & \checkmark & \checkmark &                           &                           &                           & \checkmark &                                    \\
                            & \cite{worsham_multi_2020}       &                                  &                                 &                                  &                                  & \checkmark &                           &                           & \checkmark &                           &                           &                           &                           &                                    \\ \hline
\multirow{2}{*}{MTL-RSs}    & \cite{wang2023multi}              &                                  & \checkmark       &                                  &                                  & \checkmark & \checkmark & \checkmark & \checkmark &                           & \checkmark &                           & \checkmark &                                    \\
                            & Ours                                               & \checkmark        & \checkmark       & \checkmark        & \checkmark        & \checkmark & \checkmark & \checkmark & \checkmark & \checkmark & \checkmark & \checkmark & \checkmark & \checkmark          \\ \hline
\end{tabular}
}
\begin{tablenotes}
    \item Note: CDR: cross-domain RSs, IRSs: interactive RSs, CRs: Conversational RSs, ERS : explainable RSs
    \item GNN: graph neural network, RL: reinforcement learning, KG: knowledge graph.
\end{tablenotes}
\end{table*}

\textbf{Contribution of this survey.} The goal of this survey is to comprehensively review the literature on recent advancements in MTRSs, providing a valuable reference for researchers, practitioners, and educators who wish to quickly grasp and explore the field of MTRSs. We hope that this survey will serve as practice guide for those facing the challenge of selecting the most appropriate MTL approach for solving recommendation tasks. Next, the major contributions of this survey can be summarized as follows: 

\begin{itemize}
    \item \textbf{Comprehensive Review}: We provide a comprehensive review of state-of-the-art MTRSs RSs, offering a holistic perspective on the field's current progress and challenges.
    \item \textbf{Structured Categorization}: We propose a structured categorization of existing works based on task relations, model architectures, optimization strategies, training techniques, and applications, enabling clear differentiation of MTRSs recommendation approaches.
    \item \textbf{Bridging Academia and Industry}: This survey highlights the interplay between academic advancements and industrial applications, showcasing the theoretical and practical value of MTL in RSs.
    \item \textbf{Future Direction}: We identify existing gaps and challenges in MTRSs RSs, providing actionable future directions and a roadmap for emerging researchers.
\end{itemize}


The remainder of this article is organized as follows: Section \ref{sec2} introduces the backgrounds, motivation of MTRSs and our taxonomy. Then we present task relation discovery methods in Section \ref{sec3} and the architecture of MTRSs in Section \ref{sec4}. Section \ref{sec5} gives a detailed introduction of optimization strategy. Section \ref{sec6} summarizes other learning paradigms in MTRSs. Section \ref{sec7} discusses some application problems and Section \ref{sec8} gives a overview of common datasets and evaluation metric. Section \ref{sec9} discusses the challenges and prominent open research issues. Section \ref{sec10} concludes the paper.

\section{Background and Preliminaries}\label{sec2}

Before delving into the core of the survey, we introduce the MTL technique, and present the motivation of applying MTL in RSs. Then, we also describe the definition and formalization of MTL recommendation. Additionally, we discuss the categories of MTRSs.

\subsection{Problem Formulation of Multi-Task Recommendation}

In this section, we provide a formal definition of MTRSs and present a general mathematical formulation to unify various architectures discussed in subsequent sections.

\begin{rmk}[Multi-Task RSs]
Unlike traditional single-task approaches that optimize distinct objectives independently, \textbf{MTRSs} aim to simultaneously optimize multiple related objectives (e.g., CTR and CVR) within a unified framework. By leveraging shared latent representations and modeling task correlations, MTRSs seek to improve the generalization performance of all tasks, particularly for data-sparse auxiliary tasks.
\end{rmk}

Formally, suppose we have a set of $K$ related tasks $\mathcal{T} = \{1, 2, \ldots, K\}$. The goal of MTRS is to learn a function that predicts the labels for all $K$ tasks simultaneously given an input instance.

Let $\mathcal{D} = \{(x_i, y_i^{(1)}, \ldots, y_i^{(K)})\}_{i=1}^N$ denote the training dataset with $N$ samples. For the $i$-th sample, $x_i$ represents the input feature vector (containing user profile $u_i$, item attributes $v_i$, and context $c_i$), while $y_i^{(k)}$ denotes the ground-truth label for the $k$-th task (e.g., click or purchase). The general formulation of MTRS consists of three key phases: shared feature extraction, task-specific processing, and multi-objective optimization.

\paragraph{Shared and Task-Specific Representations}
The core hypothesis of MTL is that related tasks share common underlying patterns. We denote the shared feature encoder as $f_s(\cdot; \theta_s)$, parameterized by $\theta_s$. This encoder maps the input $x_i$ into a shared latent representation $h_s$:
\begin{equation}
    h_{s,i} = f_s(x_i; \theta_s)
\end{equation}
To capture the unique characteristics of each task, MTRSs typically employ task-specific networks $f_k(\cdot; \theta_k)$ for each task $k \in \mathcal{T}$. The task-specific representation $h_{k,i}$ is derived from the shared representation (and potentially task-specific inputs):
\begin{equation}
    h_{k,i} = f_k(h_{s,i}; \theta_k), \quad k = 1, \ldots, K
\end{equation}
where $\theta_k$ represents the parameters specific to the $k$-th task.

\paragraph{Prediction and Optimization}
Based on the task-specific representation $h_{k,i}$, a prediction head (e.g., a sigmoid layer for binary classification) generates the estimated score $\hat{y}_i^{(k)}$:
\begin{equation}
    \hat{y}_i^{(k)} = g_k(h_{k,i}; \phi_k)
\end{equation}
where $\phi_k$ denotes the parameters of the prediction head. 

To train the model, we define a loss function $\mathcal{L}_k(\cdot, \cdot)$ for each task (e.g., Log-Loss for classification or MSE for regression). The final objective is to minimize the weighted sum of losses across all tasks:
\begin{equation}
    \mathcal{L}_{\text{total}} = \sum_{i=1}^N \sum_{k=1}^K \lambda_k \mathcal{L}_k(y_i^{(k)}, \hat{y}_i^{(k)}) + \Omega(\Theta)
\end{equation}
where $\lambda_k$ is a hyperparameter balancing the importance of task $k$, $\Theta = \{\theta_s, \theta_1, \ldots, \theta_K\}$ denotes the unified parameter set, and $\Omega(\Theta)$ is the regularization term.

\subsection{Extensions and Variants of Multi-Task Recommendation}

With the rapid development of RSs, user interests and behaviors have become increasingly complex, often exhibiting multi-dimensional and diverse patterns. In recent years, researchers have begun to focus on emerging problems such as multi-behavior recommendation, multi-scenario recommendation, and multi-domain recommendation. These problems extend the definition of recommendation tasks from different perspectives and further broaden the research scope of MTRSs.

\begin{figure*}
	\centering
	\subfigure[MBR]{
		\begin{minipage}[t]{0.33\linewidth}
			\centering
			\includegraphics[width=2.0in]{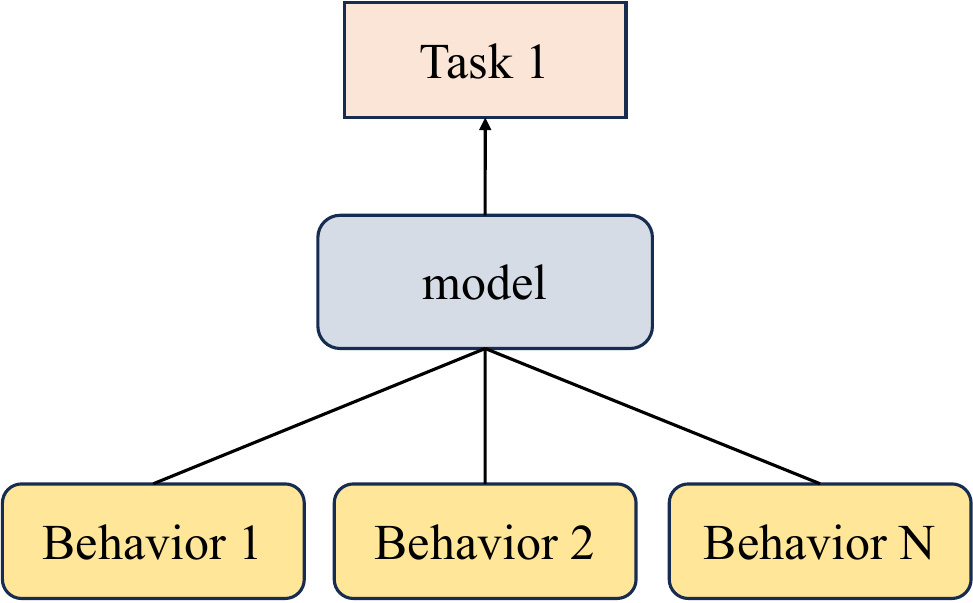}
		\end{minipage}
	}%
	\subfigure[MSR]{
		\begin{minipage}[t]{0.33\linewidth}
			\centering
			\includegraphics[width=2.0in]{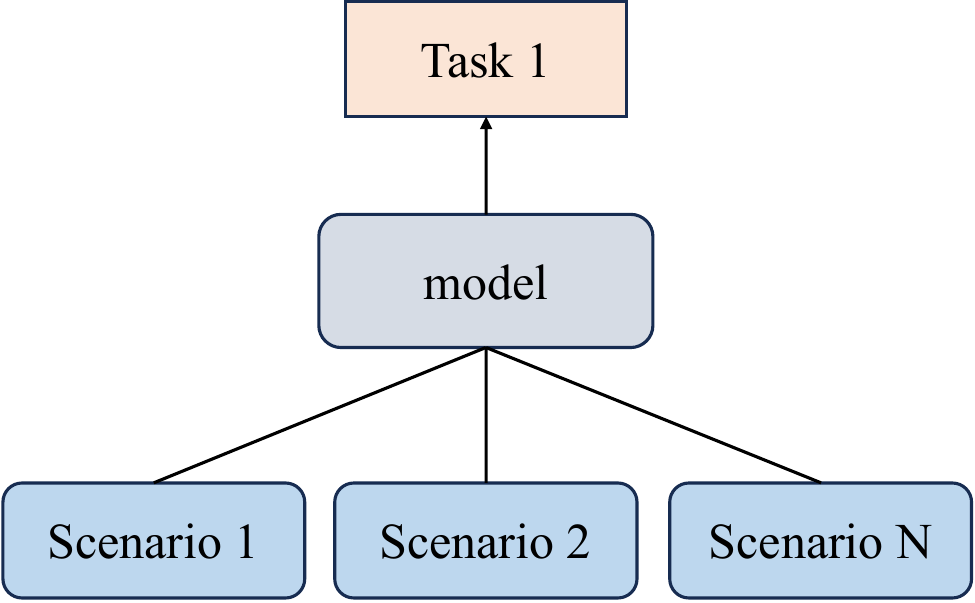}
		\end{minipage}
	}%
	\subfigure[MDR]{
		\begin{minipage}[t]{0.33\linewidth}
			\centering
			\includegraphics[width=2.0in]{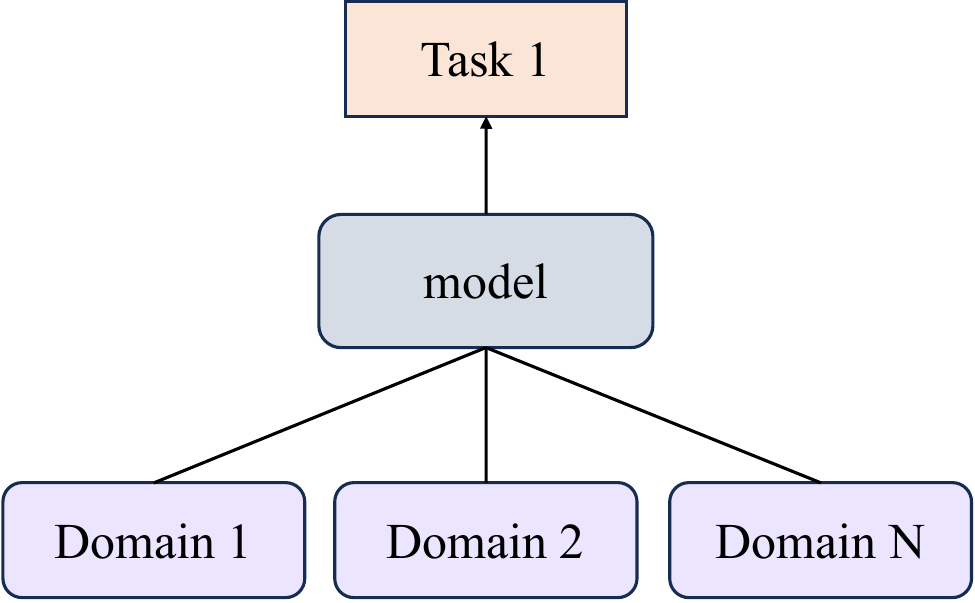}
		\end{minipage}
	}%
    
	\subfigure[Both MBR \& MTL]{
		\begin{minipage}[t]{0.33\linewidth}
			\centering
			\includegraphics[width=2.0in]{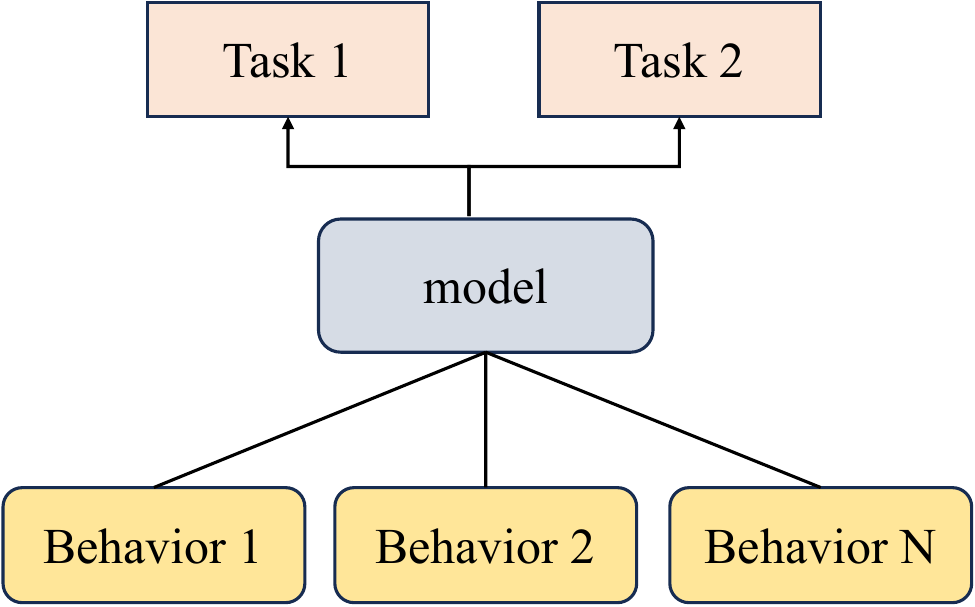}
		\end{minipage}
	}%
	\subfigure[Both MSR \& MTL]{
		\begin{minipage}[t]{0.33\linewidth}
			\centering
			\includegraphics[width=2.0in]{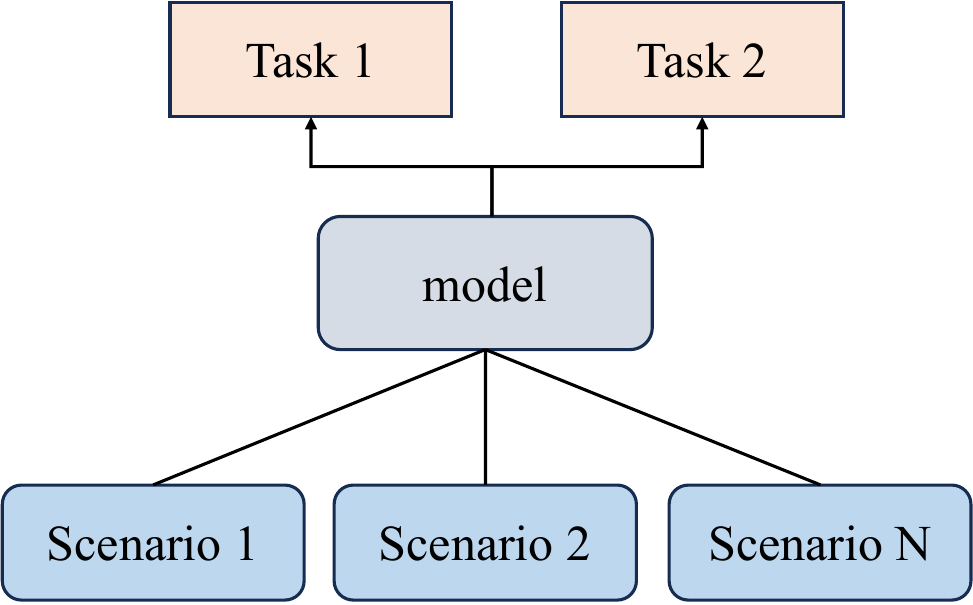}
		\end{minipage}
	}%
	\subfigure[Both MDR \& MTL]{
		\begin{minipage}[t]{0.33\linewidth}
			\centering
			\includegraphics[width=2.0in]{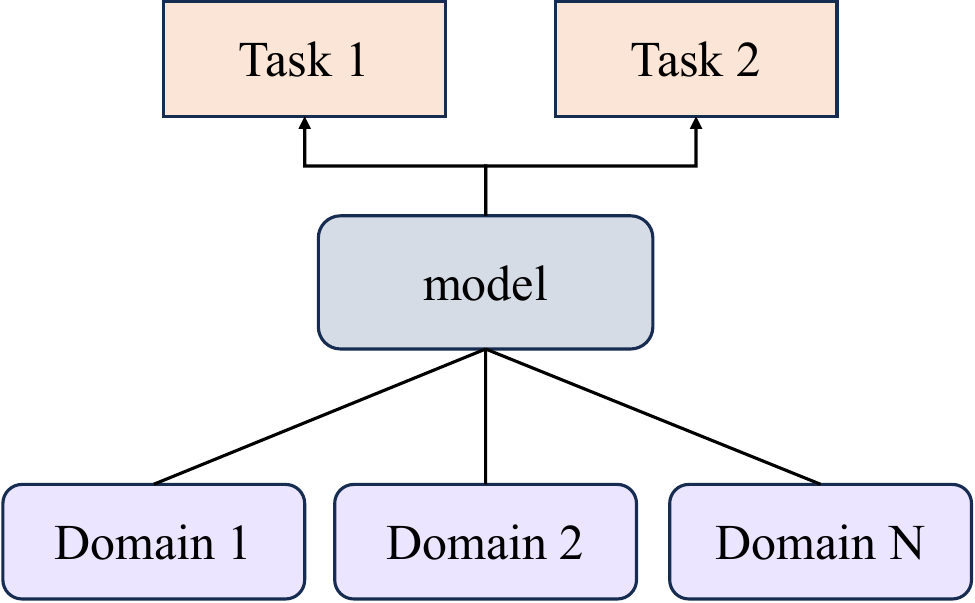}
		\end{minipage}
	}%
	\centering
	\caption{Multi-behavior, Multi-scenario, Multi-domain recommendation}
	\label{extention}
\end{figure*}

\subsubsection{Multi-behavior Recommendation}

Multi-behavior recommendation(MBR), as shown in Fig\ref{extention} (a) focuses on modeling various types of user interactions with items, such as click, add-to-cart, favorite, and purchase. Different behaviors often reflect different stages and intensities of user interest. Modeling the correlations among these behaviors helps improve the prediction of key behavior (e.g., purchase) and alleviates data sparsity problem.

In this area, existing research primarily adopts two mainstream modeling approaches:

\begin{itemize}
    \item \textbf{Sequential modeling of multi-behavior data}

 This approach integrates different user behaviors into a unified temporal sequence and employs sequential models (e.g., RNNs, Transformer) to capture the dynamic dependencies among behaviors, thereby modeling the evolution of user preferences. These methods typically take behavior sequences as input and utilize behavior-type embeddings or fusion mechanisms to distinguish between behavior types. Representative work includes MBHT \cite{yang2022multi}, PBAT\cite{su2023personalized}.
    \item \textbf{MTL based on inter-behavior relationships}

 In this approach, different behaviors are treated as related but distinct prediction tasks. MTL frameworks are applied to jointly model shared and task-specific knowledge across behaviors. A common practice is to share the bottom-layer user/item representations while employing task-specific branches, gating mechanisms, or expert networks at higher levels to capture behavioral differences. Representative methods include MMoE, PLE, and similar architectures.
\end{itemize}

\subsubsection{Multi-scenario Recommendation}

Multi-scenario recommendation(MSR), as shown in Fig \ref{extention} (b) emphasizes the differences in user preferences across various usage scenarios, such as homepage recommendation, search result pages, and item detail pages. Each scenario may differ significantly in terms of information presentation, interaction patterns, and user intent. By treating different scenarios as distinct recommendation tasks, the system can better capture users’ scenario-specific preferences, while enabling cross-scenario collaboration through shared user representations. A detailed discussion on the integration of multi-scenario and MTRSs is provided in Section \ref{msr}, as illustrated in Fig \ref{extention} (e).

\subsubsection{Multi-domain Recommendation}

Multi-domain recommendation(MDR) aims to model user preference transfer across different content domains (e.g., books, movies, and e-commerce products) or platforms, as shown in Fig\ref{extention} (c). For overlapping users, the item spaces across domains are often different. By leveraging MTL frameworks, it is possible to share user preferences while simultaneously learning domain-specific item representations, thereby improving the effectiveness of cross-domain recommendation. A detailed discussion on the integration of multi-domain and multi-task recommendation is provided in Section \ref{mdr}, as illustrated in Fig \ref{extention} (f).

\subsection{Why Apply MTL in RSs}

The primary objective of RSs  is to accurately capture user preferences and deliver high-quality content. However, traditional single-task models, relying on solitary supervision signals (e.g., clicks), often fail to represent diverse user intents and are vulnerable to data sparsity, selection bias, and cold-start problems. MTL offers a principled solution by jointly learning multiple correlated objectives. In this subsection, we elucidate why MTL is intrinsically suitable for RSs from three complementary perspectives: \textit{theoretical foundations}, \textit{empirical evidence}, and \textit{alignment with recommendation challenges}.

\paragraph{(1) Theoretical Motivation}
The efficacy of MTL in RSs is grounded in robust learning theory. Classical analyses posit that learning multiple related tasks simultaneously introduces an \textit{inductive bias}, which constrains the hypothesis space and reduces generalization error \cite{mitchell_multitask_1997, baxter2000model}. Theoretically, auxiliary tasks function as regularizers, providing supplementary perspectives that prevent the model from overfitting to noisy single-task data \cite{ruder_overview_2017}. Furthermore, multi-task risk minimization theory demonstrates that combining related tasks reduces estimation variance and enhances robustness \cite{ciliberto2017consistent, argyriou2008convex}. These insights suggest that MTL is not merely an engineering trick but a theoretically sound approach to refining feature representations, making it particularly effective for inferring preferences from sparse or biased interaction data.

\paragraph{(2) Empirical Evidence}
Extensive empirical studies validate the practical superiority of MTL in two key dimensions: algorithmic performance and business impact.

\textit{Offline Methodological Validation:} MTL architectures have proven effective in resolving specific algorithmic bottlenecks. For instance, ESMM~\cite{ma_entire_2018} demonstrates that jointly modeling CTR and CVR over the entire impression space effectively eliminates Sample Selection Bias (SSB), yielding a 2.18\% AUC gain on CVR tasks. Similarly, PLE~\cite{tang_progressive_2020} shows consistent improvements on conflicting tasks (e.g., View-Through Rate vs. View Completion Rate) by structurally disentangling shared and specific components.

\textit{Online Business Impact:} Beyond offline metrics, MTL has delivered tangible outcomes in large-scale industrial systems. In Google’s recommendation scenario, MMoE~\cite{ma_modeling_2018} achieved a 0.25\% lift in engagement and a 2.65\% improvement in satisfaction. In FinTech scenarios, ESCM\(^2\) deployed by Ant Insurance recorded substantial lifts of 5.64\% and 10.85\% in CVRs and insurance premiums, respectively. Furthermore, STEM-Net~\cite{su2024stem} achieved an average AUC lift of roughly 0.35\% across multiple scenarios, translating to a significant GMV growth of up to 7.1\% in Tencent's advertising platform. These results confirm that utilizing diverse behavioral signals (clicks, likes, purchases) via MTL leads to richer preference representations and substantial commercial gains.

\paragraph{(3) Theoretical Alignment with Core Challenges}
Finally, the mechanisms of MTL naturally align with the intrinsic bottlenecks of RSs. \textbf{Data sparsity} is ameliorated by transferring knowledge from dense auxiliary tasks (e.g., clicks) to sparse target tasks (e.g., purchases). \textbf{Selection bias} is mitigated by modeling the sequential dependency of user behaviors over the entire space. \textbf{System efficiency} is optimized by consolidating disjoint pipelines into a unified network. This structural alignment ensures that MTL is not just a performance booster, but a strategic necessity for modern recommendation engines.

Overall, supported by theoretical guarantees, empirical success, and problem-solution alignment, MTL has established itself as a foundational paradigm in the recommendation community.

\subsection{Taxonomy of MTRSs}

Through sorting out the literature and review the motivation of MTRSs, we conclude three challenges as follows: 

\begin{itemize}
    \item \textit{\textbf{Challenge 1}}. How to identify which tasks can be jointly learned to enhance the performance of each individual task?
    \item \textit{\textbf{Challenge 2}}. How to design an effective mechanism to learn common features across tasks?
    \item \textit{\textbf{Challenge 3}}. How to balance task-specific objectives during multi-task optimization and ensure stable and effective training?
\end{itemize}

\begin{figure}
    \centering
    \includegraphics[width=1\linewidth]{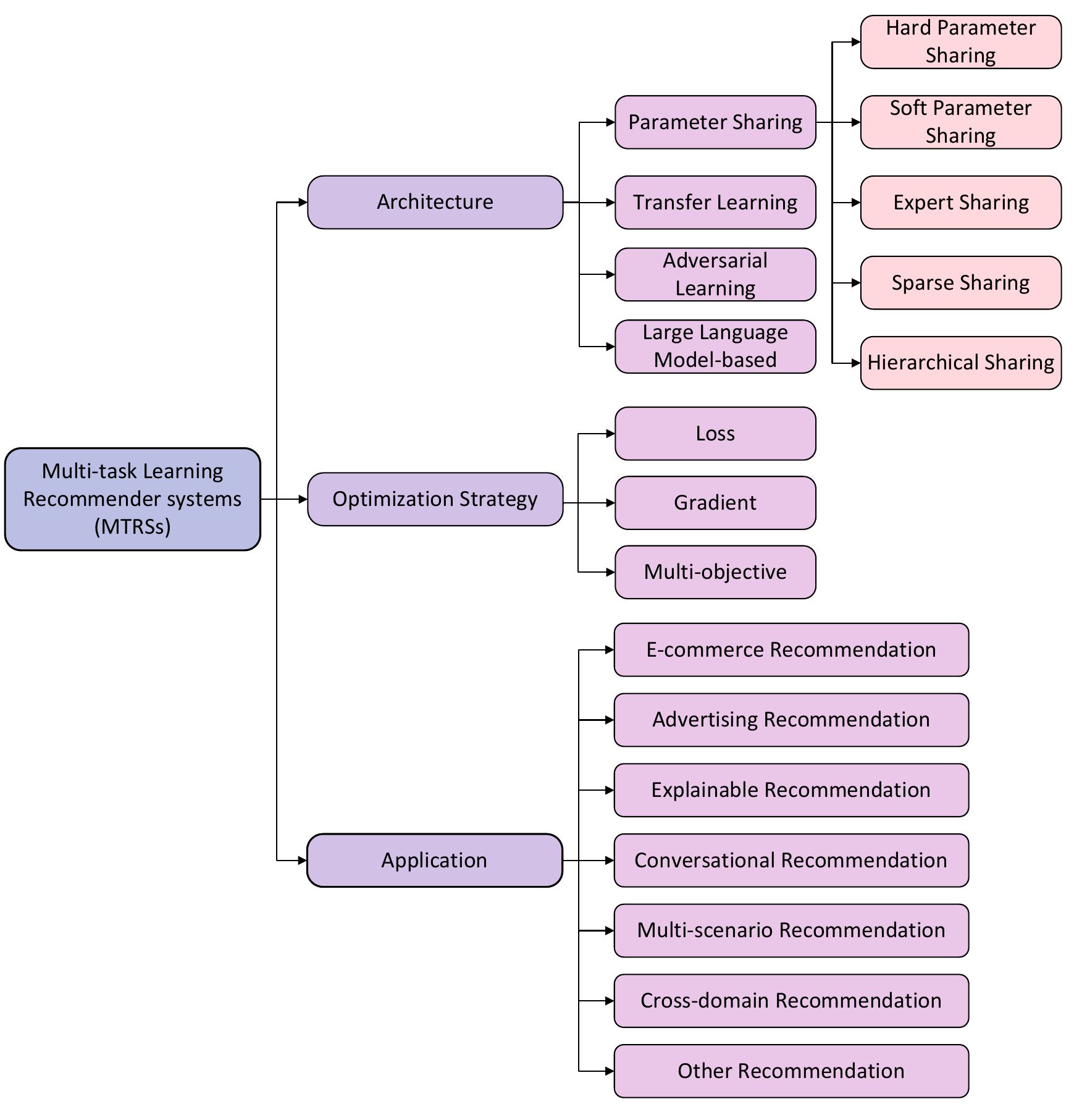}
    \caption{Categories of MTRSs}
    \label{fig2}
\end{figure}

The most recent works aim to propose efficient techniques to solve above three challenges, so we categorize them into three corresponding taxonomies: task relation discovery, the architecture of MTRSs, optimization and training strategies, as shown in Figure \ref{fig2}. The first category focuses on \textit{\textbf{Challenge 1}}, which aim to explore the relationship between tasks, separate counterproductive tasks, and put mutually reinforcing tasks on the same network for learning as much as possible. The researches on MTRSs model structure is currently a top priority in academia and industry, which design appropriate shared layers and task-specific layers to improve the tasks performance. It refers to \textit{\textbf{Challenge 2}}. To handle \textit{\textbf{Challenge 3}}, optimization strategies focus on balancing task-specific losses through approaches such as adaptive weighting and resolving gradient conflicts during the optimization process. Additionally, training paradigms—including joint learning, reinforcement learning, and pre-training with fine-tuning—are designed to enhance the efficiency and effectiveness of MTL while harmonizing task collaboration and independence. Furthermore, beyond technical challenges, we highlight the importance of applying MTL to real-world RSs, discussing current application challenges and potential directions for improvement.

\section{Task Relationship in MTRSs}\label{sec3}

The effectiveness of MTRSs critically depends on how information is organized, shared, and differentiated across tasks, and such behaviors are fundamentally shaped by the underlying task relationships. Although prior studies widely acknowledge the importance of task relationships, the literature lacks a unified and systematic framework that characterizes relationship types, analyzes their effects on training dynamics and cross-task information flow, and reveals the potential challenges such as task conflicts and negative transfer. This section introduces a two-dimensional taxonomy of recommendation task relationships based on explicit dependency structures and the presence of auxiliary tasks, and explains why task relationships play a pivotal role in shaping performance in MTRSs. We then provide an in-depth discussion of task conflicts and negative transfer, covering the causes, manifestations, detection techniques, and mitigation strategies. Together, these analyses establish the conceptual foundation for model architecture design and optimization strategies in subsequent sections.

\subsection{A Two-Dimensional Taxonomy of Task Relationships}

Task relationships serve as the structural foundation for information sharing and representation learning in MTRSs. To systematically characterize the diverse interaction patterns observed in existing MTRSs, we develop a two-dimensional taxonomy based on explicit dependency structures and auxiliary-task design, as shown in Table \ref{tab_task}. The first dimension focuses on capturing how tasks are intrinsically related during joint learning, while the latter concerns whether auxiliary objectives are introduced to enhance representation learning, improve generalization. This subsection elaborates on these dimensions and their combinations, supported by representative examples from MTRSs.

\begin{table*}[ht]
\centering
\caption{A Two-Dimensional Taxonomy of Task Relationships in MTRSs}
\label{tab_task}
\begin{tabular}{p{0.12\linewidth} p{0.24\linewidth} p{0.24\linewidth} p{0.28\linewidth}}
\hline
\textbf{Task Relation} & 
Without Auxiliary Tasks & 
With Auxiliary Tasks & 
Representative Works \\
\hline

Independent Parallel Tasks &
Joint optimization of independent binary classification objectives (e.g., CTR \& CVR prediction; multi-behavior prediction). &
Integration of self-supervised learning objectives (e.g., contrastive loss) for robust representation learning. &
MMoE\cite{ma_modeling_2018}, DUPN\cite{ni_perceive_2018}, PLE\cite{tang_progressive_2020}, MSSM\cite{ding_mssm_2021}, Contrastive MTL\cite{bai_contrastive_2022}, PEPNet\cite{chang_pepnet_2023}, MoME\cite{xu2024mome}, M\(^3\)oE\cite{zhang2024m3oe} \\

Sequential Dependency Tasks &
Modeling sequential conversion funnel with probabilistic dependencies &
Incorporating auxiliary objectives to mitigate data sparsity in downstream tasks &
ESMM\cite{ma_entire_2018}, ESM\(^2\)\cite{wen_entire_2020}, ESCM$^{2}$\cite{wang_escm2_2022}, HM$^{3}$ \cite{wen_hierarchically_2021}, ESDF\cite{wang2020delayed}, AITM\cite{xi_modeling_2021}, ECAD\cite{zhao2023entire}, AEM$^{2}$TL\cite{yi2025adaptive} \\

\hline
\end{tabular}
\end{table*}

\subsubsection{Parallel Tasks}
Parallel task relationships exist when multiple tasks are optimized simultaneously, and there is no explicit sequential or causal dependency between their target behaviors. In this setting, tasks typically share the bottom layers of the model to learn common underlying representations, while task-specific top layers (towers) are employed to optimize their respective objectives independently. The overall objective function is usually a weighted sum of individual task losses. This structure is widely used when tasks are related but hold equal status, such as predicting CTR and CVR tasks concurrently, or modeling diverse user behaviors (e.g., clicking, liking, favoring) in parallel.

\paragraph{\textbf{Base Models Without Explicit Auxiliary Tasks}}
Early research focused on designing architectures to effectively share information across parallel tasks without explicitly introducing auxiliary objectives. The foundational work, Shared-Bottom \cite{mitchell_multitask_1997}, established the paradigm of sharing bottom layers while keeping top layers task-specific. While efficient, it often suffers from negative transfer due to task conflicts. To address this, MMOE \cite{ma_modeling_2018} introduced Multi-gate Mixture-of-Experts to dynamically learn task relationships and route information. PLE \cite{tang_progressive_2020} further improved upon MMOE by explicitly separating shared and task-specific experts to mitigate the "seesaw phenomenon." Building on these foundations, recent works have focused on more dynamic and personalized parameter sharing mechanisms. For instance, PEPNet \cite{chang_pepnet_2023} introduces Parameter and Embedding Personalized Gate networks to adaptively fuse features based on input-specific information across tasks. Similarly, M3oE \cite{zhang2024m3oe} proposes a multi-horizon, multi-objective architecture that utilizes mixture-of-experts to capture complex task correlations in diverse recommendation scenarios. These architectures form the backbone of many modern parallel MTL applications.

Several works apply this parallel paradigm to specific recommendation scenarios. For instance, DUPN \cite{ni_perceive_2018} learns universal user representations across multiple e-commerce tasks. In the context of multi-behavior recommendation, MSSM \cite{ding_mssm_2021} models different types of user-item interactions as parallel tasks.

\paragraph{\textbf{Enhanced Models With Auxiliary Tasks}}
While base models focus on architecture, recent advancements often integrate auxiliary tasks to introduce additional supervision, regularization, or knowledge. A prominent trend is using Self-Supervised Learning (SSL) to learn more robust representations. For instance, Contrastive MTL \cite{bai_contrastive_2022} introduces contrastive learning to regularize primary tasks by maximizing mutual information between different views. Further advancing this direction in multi-behavior scenarios, KMCLR \cite{xuan2023knowledge} proposes two Contrastive Learning tasks as an auxiliary task to model coarse-grained commonalities between behaviors and fine-grained differences between users, thereby improving the recommendation performances. Beyond SSL, researchers have explored knowledge distillation and causal inference to enhance model learning. CrossDistil \cite{yang_cross-task_2022} proposes cross-task knowledge distillation, where tasks act as teachers for each other, while CFS-MTL \cite{chen_cfs-mtl_2022} selects stable causal features from a causal perspective as an auxiliary task to enhance stability against data biases. Additionally, some methods leverage auxiliary tasks calibrated with small amounts of unbiased data to correct biases in the primary tasks. For example, \textbf{Res-DR} \cite{li2023removing} introduces a consistency loss based on unbiased data to calibrate the primary recommendation task.

\subsubsection{Sequential Dependency Tasks}
In contrast to parallel tasks, sequential dependency tasks exhibit a clear chronological or causal order, where the occurrence of one task's target behavior is conditional on the preceding one (e.g., Exposure $\to$ Click $\to$ Conversion). In this structure, information flow is inherently one-way or cascaded, and models must respect this dependency to avoid issues like Sample Selection Bias (SSB) and data sparsity in downstream tasks.

\paragraph{\textbf{Cascaded Architectures for Sequential Modeling}}
The primary goal of models in this category is to leverage the sequential relationship, often using data-rich upstream tasks to assist data-sparse downstream tasks. The seminal work, \textbf{ESMM} \cite{ma_entire_2018}, proposed an Entire Space Multi-task Model to simultaneously model CTR and CVR, implicitly treating the dense CTR task as auxiliary supervision for the sparse CVR task to address SSB. \textbf{ESM$^2$} \cite{wen_entire_2020} extends the approach to handle delayed feedback. More advanced architectures have been proposed to better model the information transfer across the sequence. \textbf{AITM} \cite{xi_modeling_2021} introduces an Adaptive Information Transfer mechanism via attention. Recently, \textbf{AEM$^2$TL} \cite{yi2025adaptive} proposes an adaptive entire-space model, extending sequential modeling to complex multi-scenario settings.

\paragraph{\textbf{Incorporating Additional Auxiliary Objectives}}
Beyond the inherent auxiliary role of upstream tasks, researchers have introduced explicit additional auxiliary objectives to further enhance modeling of specific stages or challenges in the sequence. To bridge the gap between main sequential stages, intermediate and micro-behavior modeling has been explored. For instance, HM$^3$ \cite{wen_hierarchically_2021} models micro-behaviors (e.g., dwelling, scrolling) as intermediate auxiliary tasks. Similarly, works like MB-GMN \cite{xia2021graph} utilize graph neural networks to model the cascading relationship between diverse multi-behavior patterns as auxiliary signals for the target behavior. Furthermore, causal and counterfactual learning has been applied to address specific biases; ESCM$^2$ \cite{wang_escm2_2022} employs counterfactual risk minimization as an auxiliary objective to address biases in CVR estimation within the sequential framework. To address the issue of delayed feedback in real-world streams, ESDF \cite{wang2020delayed} introduces auxiliary mechanisms to model the time delay between clicks and conversions. More recently, Finally, representation reshaping techniques have been proposed to optimize information transfer. AEM$^{2}$TL\cite{yi2025adaptive} leverages conditional probability rules derived from the post-impression behavior graph to jointly model auxiliary tasks and CVR prediction within a unified multi-scenario and multi-task learning framework.

\subsection{Why Task Relationship Matters in Recommendation}

In MTRSs, whether tasks have exploitable relationships directly determines whether shared representations will yield performance gains. Classic MTL theory highlights that the effectiveness of parameter sharing is highly dependent on the relatedness between tasks. When tasks are unrelated or even contradictory, shared structures can lead to the learning of suboptimal representations \cite{mitchell_multitask_1997, ruder_sluice_2017, ruder_overview_2017}. In recommendation systems, this issue presents unique complexities. While most tasks (such as CTR, CVR and dwell time) stem from the same user's behavior and thus have inherent correlations, these relationships are not simply linear positive correlations; they exhibit a coexistence of dependence and competition. For example, there is an explicit sequential dependency between CTR and CVR: conversion must necessarily occur after a click. However, the relationship between CTR and dwell time may exhibit misalignment, where content with misleading titles may gain high clicks but result in negative feedback due to low dwell time \cite{yi2014beyond, covington2016deep}. Thus, in recommendation settings, the challenge of MTL lies not in whether tasks are related, but in how to leverage the relationships between tasks and coordinate these heterogeneous objectives within a unified framework, avoiding negative transfer caused by inconsistent gradients or differences in data distribution (e.g., sparsity).

The relationships between different tasks directly influence the design of multi-task recommendation models. For example, the strong conditional dependence between CTR and CVR has led to the widespread use of cascade models, such as ESMM \cite{ma_entire_2018} and AITM \cite{xi_modeling_2021}, which explicitly model the conversion process to enhance performance. In contrast, for parallel tasks (such as CTR, add-to-cart (ATC), and save-to-favorite), models typically employ shared backbone networks with separate tower structures. For tasks with hierarchical dependencies, such as micro-behaviors and macro-behaviors \cite{wen_hierarchically_2021}, hierarchical structures are required to explicitly model behavior sequences. Therefore, task relationships determine the mode of parameter sharing and drive the development of specialized model architectures.

If task relationships are ignored and blind sharing is adopted, task conflicts can lead to negative transfer. For instance, in recommendation systems, the data distributions between stages such as "exposure → click → conversion" differ significantly. Exposure behavior is influenced by position bias, while purchase behavior contains long-tail sparsity, and these two tasks often have inconsistent gradient directions. If shared directly, the patterns can become incorrectly mixed in the unified representation space, disrupting the learning of key features. As a result, numerous studies have proposed strategies for different sharing architecture (in Section \ref{sec4}), or resolving gradient conflicts (in Section \ref{sec5}). Therefore, improper handling of task relationships is a core cause of performance degradation in multi-task recommendation systems.

\subsection{Task Conflicts and Negative Transfer}

\paragraph{The Phenomenon of Negative Transfer}  
Despite the intention of MTL to achieve mutual benefit through shared representations, negative transfer often occurs in practice, where the performance of multi-task models is actually worse than that of models trained independently on individual tasks \cite{tang_progressive_2020}. This performance degradation is primarily due to fundamental conflicts during the optimization process, which prevent the shared parameters from simultaneously satisfying the optimal solution requirements for all tasks. In recommendation systems, these conflicts are primarily manifested in two mechanisms: gradient direction conflict and magnitude dominance, which will be discussed below.

\paragraph{Gradient Direction Conflict}  
Gradient direction conflict represents a direct mathematical manifestation of task goal inconsistency. During backpropagation, each task \( k \) produces a gradient vector \( g_k \) for the shared parameters \( \theta_s \). When tasks are in competition (e.g., inducing clicks vs. user satisfaction), their gradient vectors may exhibit negative cosine similarity or even be opposite in direction. This phenomenon is known as “gradient interference” or “cancellation interference”:
$$
\cos(g_i, g_j) < 0
$$
Physically, this means one task tries to push parameters “to the left,” while another task tries to push them “to the right.” As a result, the shared parameters oscillate or stagnate on the optimization surface, preventing the model from converging to a region that benefits all tasks simultaneously \cite{yu2020gradient}. This conflict makes it harder for multi-task models to converge to an optimal solution that serves all tasks effectively.

\paragraph{Magnitude Dominance and Data Imbalance}  
Even when gradient directions align, differences in gradient magnitude can still lead to negative transfer. This issue is particularly prominent in recommendation systems, where the training data for different tasks often differ by several orders of magnitude. For instance, CTR task typically has millions of positive samples, while CVR or purchase task may only have tens of thousands. This disparity in data sparsity leads to significantly larger gradient magnitudes for the CTR task:
$$
||g_{CTR}|| \gg ||g_{CVR}||
$$
In the gradient update process, the gradient of the dominant (head) task can overwhelm the gradient of the tail tasks, leading to an update direction that favors the head task. This results in shared representations being overly biased toward fitting the head task, thereby sacrificing the feature expressiveness of the tail tasks. Consequently, the auxiliary tasks not only fail to help but are instead “noise” for the model.

\paragraph{Optimization Dilemma}  
In summary, the gradient direction conflicts and magnitude imbalances contribute to the optimization dilemma in multi-task recommendation systems. Without intervention, the model typically converges to a suboptimal solution, sacrificing the performance of some tasks to minimize the overall loss. This leads to the development of techniques like gradient conflict resolution algorithms (e.g., PCGrad\cite{yu2020gradient}) and decoupled architectures (e.g., PLE \cite{tang_progressive_2020}), which are designed to mitigate negative transfer and improve the overall performance of multi-task models.

\paragraph{Future Directions and Further Exploration}  
Future work should explore dynamic task relationship adjustment strategies, particularly in large-scale, cross-domain systems, where task relationships may evolve over time. Techniques like meta-learning and reinforcement learning could be used to dynamically assess task relationships and adjust parameter sharing mechanisms accordingly. Moreover, exploring automated task selection based on real-time performance could further mitigate negative transfer in evolving systems. Current solutions like multi-task transfer learning and gradient normalization techniques are a step forward, but more research is needed to fully resolve the challenges posed by task conflicts in multi-task recommendation systems.

\section{Model Architecture of MTRSs}\label{sec4}

As deep learning technology continues to evolve, various improved MTRSs are being developed. These models often incorporate novel network architectures and learning strategies to further improve performance and generalization capabilities. Based on the research and analysis, MTL-based recommendation models can be classified into three categories: parameter sharing architecture, transfer learning, and adversarial architecture. It's important to note that there is some overlap between different architectures, and the boundaries between these architectures are not fixed and may vary depending on the specific model and task at hand. So, some MTRSs can be classified into more than one category. Below we detail MTRSs with different architectures.

\subsection{The Evolutionary Trajectory of MTRSs}

\begin{figure*}
    \centering
    \includegraphics[width=1\linewidth]{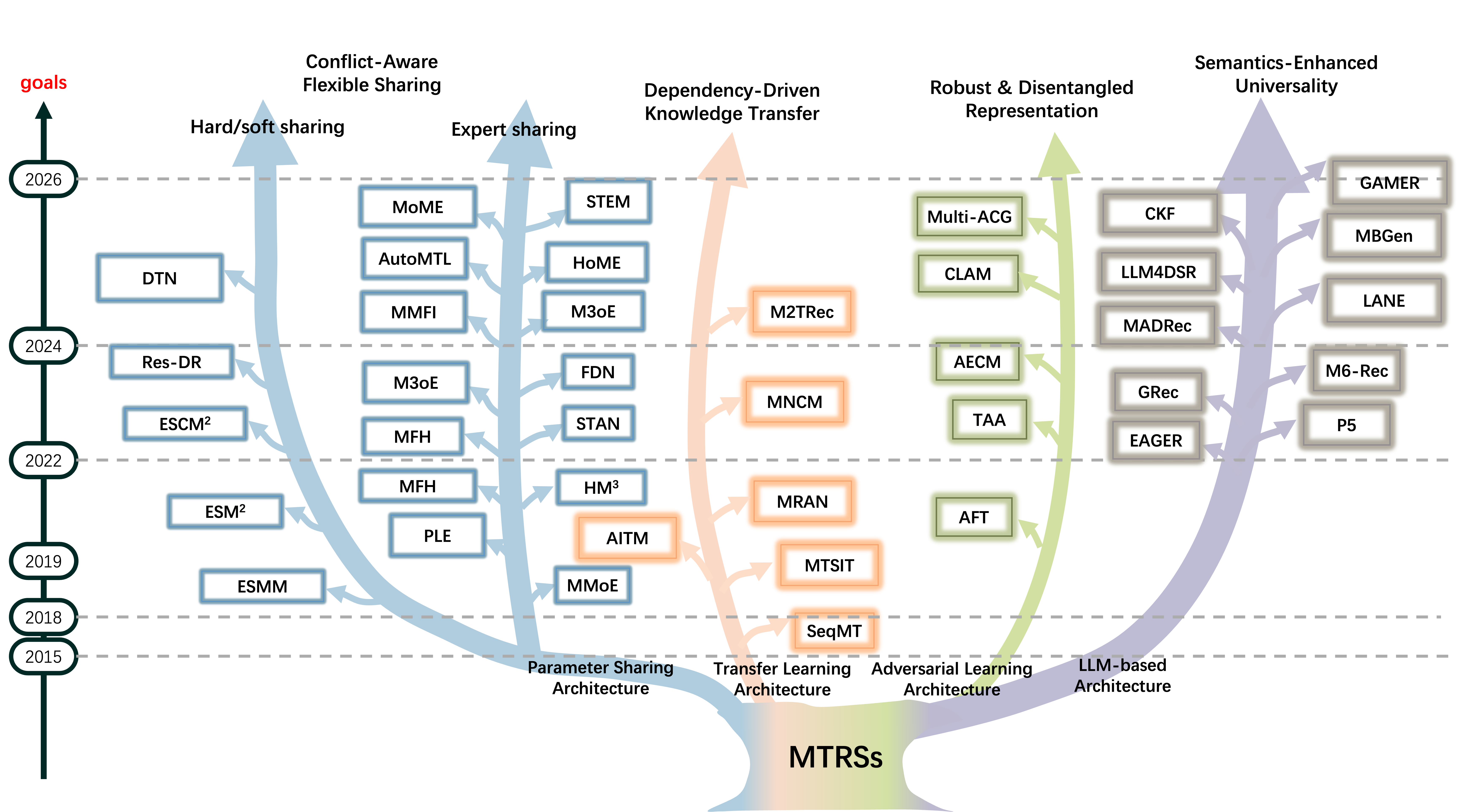}
    \caption{The Evolutionary Trajectory of MTRSs}
    \label{evolution}
\end{figure*}

The development of Multi-Task RSs is not a random collection of models but a coherent evolutionary history driven by the continuous necessity to resolve persistent industrial bottlenecks. As illustrated in Figure \ref{evolution} (we have also added an evolutionary tree figure to visualize this process), the evolutionary trajectory of this field outlines a clear progression from "efficiency-oriented rigid sharing" toward "conflict-aware flexible routing," and ultimately to a "semantics-enhanced universal paradigm."

The Hard Parameter Sharing (i.e., Shared-Bottom) paradigm, originating around 2017, constitutes the root of this evolutionary tree. Due to its superior inference efficiency and storage advantages derived from sharing substantial bottom embedding layers, it quickly became the standard choice in the industry. However, as business complexity increased, this rigid sharing mechanism exhibited severe "seesaw phenomena" (negative transfer) when facing competing tasks. This limitation became the primary driving force pushing architectural evolution.

To break this bottleneck, the field rapidly evolved toward the "Conflict-Aware Flexible Sharing" branch. Pioneering works like MMoE \cite{ma_modeling_2018} and PLE \cite{tang_progressive_2020} introduced Mixture-of-Experts (MoE) and gating mechanisms. By implementing dynamic routing to explicitly disentangle shared patterns from task-specific ones, these architectures effectively mitigated gradient conflicts, becoming the new industrial standard for multi-objective ranking.

Concurrently, to address the prevalent challenge of \textbf{data sparsity} in downstream tasks (e.g., CVR prediction), another branch—"Dependency-Driven Knowledge Transfer"—emerged. Models such as ESMM \cite{ma_entire_2018} and AITM \cite{xi_modeling_2021} abandoned parallel symmetric structures in favor of leveraging the inherent sequential dependencies of user behavior (e.g., Impression $\to$ Click $\to$ Purchase). They established asymmetric information transfer pathways from high-frequency source tasks to sparse target tasks, proving highly effective for specific sequential scenarios.

In recent years, as traditional ID-based deep recommendation models approach their performance ceilings, the field is witnessing a third significant paradigm shift. Inspired by advancements in Natural Language Processing, and aiming to bridge the semantic gap in collaborative filtering while enhancing zero-shot generalization, "LLM-based Architectures" are rapidly emerging. This nascent branch (e.g., P5 \cite{geng2022recommendation}, M6-Rec \cite{cui2022m6}) attempts to leverage the vast world knowledge and reasoning capabilities of pre-trained foundation models, reformulating recommendation tasks as sequence generation problems with the goal of achieving truly universal RSs.

In summary, the practical evolution of MTRSs reflects a continuous trade-off and optimization among efficiency, conflict resolution, sparsity mitigation, and semantic generalization.

\subsection{Parameter Sharing Architecture}

\begin{figure*}[h]
    \centering
    \includegraphics[width=0.8\linewidth]{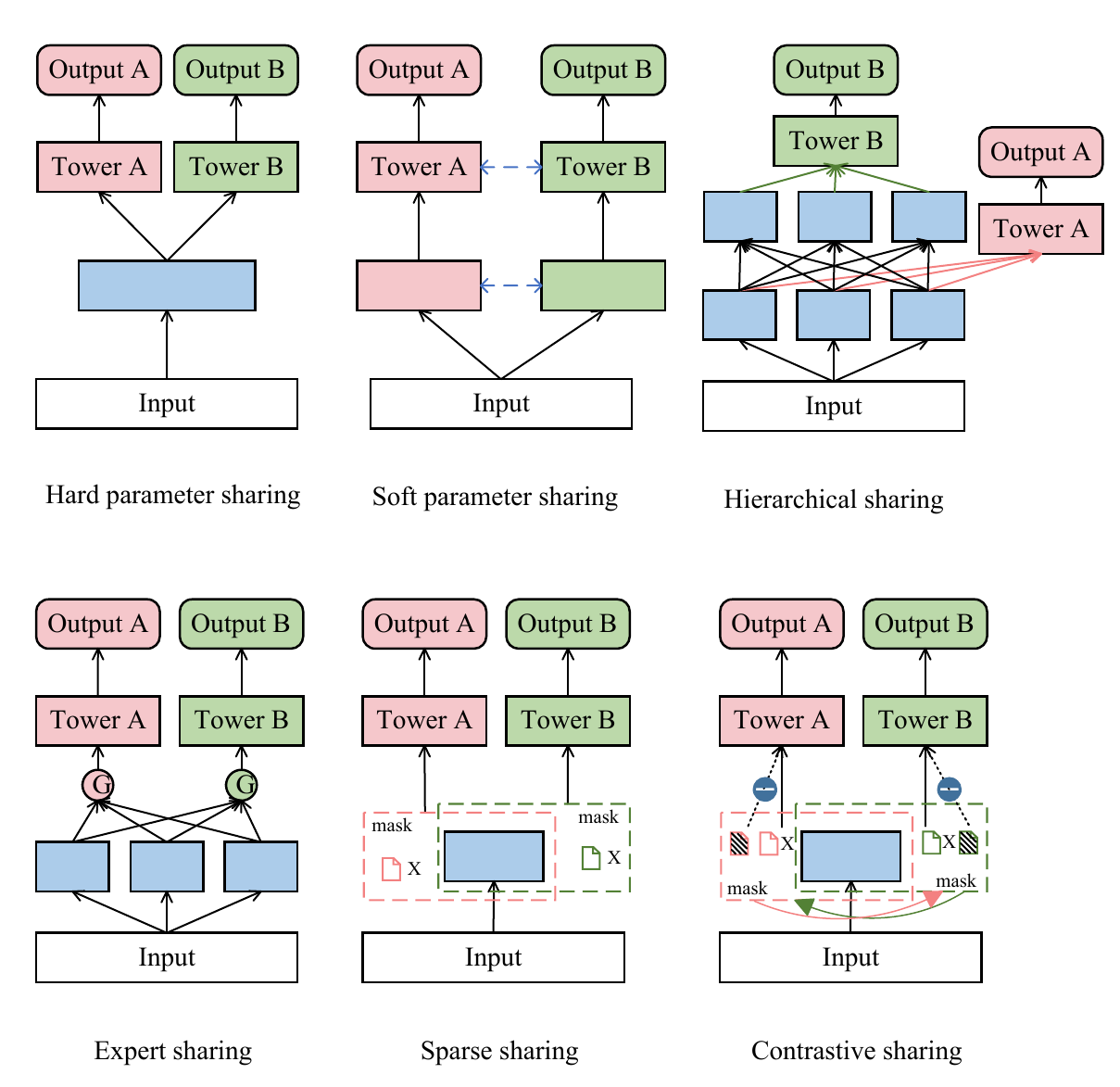}
    \caption{The illustration of hard sharing, soft sharing, expert sharing, sparse sharing, and hierarchical sharing with task A and B. Blue represents the shared parameters. pink and green represent the task-specific parameters.
    }
    \label{fig3}
\end{figure*}

One of the most widely used architectures in MTL is to share parameters in the bottom or tower layers. Much research has focused on designing effective parameter-sharing mechanisms to optimize task performance. In this section, we introduce six distinct parameter-sharing mechanisms: hard parameter sharing, soft parameter sharing, expert sharing, sparse sharing, hierarchical parameter sharing, and dynamic parameter sharing. Although sparse sharing, hierarchical sharing, and dynamic parameter sharing can often be considered subcategories of expert sharing due to their reliance on expert modules, we distinguish these mechanisms to provide a clearer explanation of their unique characteristics. Figure \ref{fig3} illustrates the model architectures corresponding to these sharing mechanisms.

\subsubsection{Hard Parameter Sharing}

Hard parameter sharing is the most widely used sharing mechanism in the previous MTL work, where multiple tasks share the same model parameters or layers, especially in the bottom layer. In \citep{Caruana_Multitask_1998}, hard parameter sharing was first proposed, and then widely used in computer vision, such as the typical target detection models R-CNN \citep{girshick_rich_2014}, YOLO \citep{redmon_you_2016}, etc. Later on, the hard parameter sharing architecture was also applied in the field of RSs. However, hard parameter sharing methods have a strong assumption, that is the tasks are strongly related and can benefit from a common representation. The key characteristics of hierarchical sharing include the following:
\begin{itemize}
    \item Common Bottom Layer: Multiple tasks share the same bottom layers, which can capture the common representations across tasks.
    \item Task-specific Tower Layers: While the bottom layers are shared, there might be task-specific layers on top of the network that are specialized for each individual task.
    \item Joint Training: All tasks are trained simultaneously on the shared layers, allowing them to benefit from shared knowledge and data.
\end{itemize}


\noindent\textbf{Existing Work.} To alleviate the data sparsity issue, many works applied the shared bottom structure of hard parameter sharing \citep{he_metabalance_2022}. For instance, ESMM \citep{ma_entire_2018, afsar_reinforcement_2022} proposes a novel approach to modeling CTR and CVR based on user behavior sequential patterns. ESM\(^2\) \citep{wen_entire_2020} extends ESMM by constructing a user sequential behavior graph and simultaneously modeling CTR prediction and auxiliary tasks based on the conditional probability rule defined on the graph. These approaches shared the bottom feature representation to learn CTR with the auxiliary task of CVR. Multi-IPW and Multi-DR \cite{zhang2020large} propose two efficient and highly effective CVR estimators to mitigate the data sparsity from a causal perspective. However, these studies only consider the selection bias induced by measured confounders. To further deal with hidden (or unmeasured) confounders, Res-IPS and Res-DR \cite{li2023removing} introduce task-specific residual networks upon the hard parameter sharing architecture. By introducing consistency loss, they utilize a small amount of unbiased data to calibrate the estimated bias, thereby addressing the issue of hidden confounding in RSs. Hard parameter sharing is a straightforward multi-task architecture that is easy to implement and is well-suited for handling tasks with strong dependencies, but it may not perform well when dealing with multiple tasks with weak dependencies.

\noindent\textbf{Advantages of Hard Parameter Sharing.}

\begin{itemize}
    \item Regularization: Sharing parameters among tasks acts as a form of regularization, which can reduce the risk of overfitting when there's limited data for one task.
    \item Memory Efficiency: Since parameters are shared among tasks, it uses less memory than using separate parameters for each task.
    \item Generalization: By forcing tasks to share a representation, it can improve the generalization of the model to new data, especially if the tasks are related.
\end{itemize}

\noindent\textbf{Limitations of Hard Parameter Sharing.}

\begin{itemize}
    \item Negative Transfer: If the tasks are not closely related, sharing parameters might lead to a situation where training on one task adversely affects the performance on another task.
    \item Less Flexibility: Since tasks share the same parameters, there's less flexibility to capture task-specific features in the shared layers.
\end{itemize}

\subsubsection{Soft Parameter Sharing}
Unlike hard parameter sharing where tasks share the exact same layers or parameters, in soft parameter sharing, tasks have distinct parameters, but these parameters are encouraged to be close to each other. The key characteristics of hierarchical sharing include the following:
\begin{itemize}
    \item Separate Models for Each Task: Each task has its own set of model parameters. However, the model architectures for each task are typically the same or similar.
    \item Regularization Between Models: A regularization term is added to the loss function to penalize large differences between the parameters of different tasks. This encourages the models of different tasks to have similar weights.
    \item Balancing Task-specific and Shared Information: The regularization term controls the balance between allowing each task to learn its own unique features and encouraging tasks to share information.
\end{itemize}

\noindent\textbf{Existing Work.} Different from hard parameter sharing methods, soft parameter parameter methods \cite{duong_low_2015} model different tasks separately and share information between tasks by calculating the relevant weights or introducing attention mechanisms. For example, Fei et al. \citet{fei_gemnn_2021} introduce a neural network-based model that allows parameter sharing from upstream tasks to downstream tasks to improve training efficiency. Chen et al. \citet{chen_co-attentive_2019} design an encoder-selector-decoder architecture for MTL and leverages multi-pointer co-attention selector to share parameter among different tasks. Chen et al. \citet{chen_cfs-mtl_2022} add feature selection modules on the top of the bottom shared network for different tasks. Xiao et al. \citet{xiao_lt4reclottery_2021} apply the neuron-connection level sharing to solve the sharing\&conflict problem in CTR and CVR, which can automatically and flexibly learn which neuron weights are shared or not shared without artificial experience. MARRS \citep{bhumika_marrs_2022} consists of three components: the wide component, the deep component (which is shared by all tasks), and the task-specific component. Each task has its own wide component and task-specific component.

\noindent\textbf{Advantages of Soft Parameter Sharing.}

\begin{itemize}
    \item Flexibility: Each task has its own parameters, allowing the model to learn task-specific representations. This can be especially beneficial when the tasks have distinct characteristics.
    \item Potential for Positive Transfer: The regularization term can guide tasks to leverage commonalities, which might lead to improved performance, especially when tasks are related.
    \item Less Risk of Negative Transfer: Since tasks have separate parameters, there's a reduced risk of negative transfer compared to hard parameter sharing. Learning in one task is less likely to adversely affect another.
\end{itemize}

\noindent\textbf{Limitations of Soft Parameter Sharing.}

\begin{itemize}
    \item Memory Overhead: Maintaining separate parameters for each task can consume more memory compared to hard parameter sharing.
    \item Regularization Coefficient: Determining the optimal regularization coefficient that dictates the extent to which parameters should be similar across tasks can be challenging and may require extensive experimentation.
    \item Complexity: Introducing a regularization term can add to the complexity of the optimization problem.

\end{itemize}

\subsubsection{Expert sharing}

Expert sharing refers to a model architecture wherein certain "experts" (sub-networks or model components) are designed to handle specific tasks or subsets of tasks. These experts can either operate individually or collectively to address a given task, allowing for both shared and task-specific representations.  The key characteristics of hierarchical sharing include the following:
\begin{itemize}
    \item Model Structure: An expert-sharing architecture might consist of multiple sub-networks (experts) alongside a gating mechanism. This gating mechanism determines which expert(s) to utilize for a given input or task.
    \item Task Specialization: Some experts might specialize in specific tasks, providing task-specific representations. Others might be more generic and can be involved in multiple tasks.
    \item Gating Mechanism: This is a critical component in expert-sharing architectures. It's responsible for determining the weight or importance of each expert's output for a given task or input. The gating mechanism can be learned alongside the main model, allowing the model to automatically determine the best expert combination for each task.
\end{itemize}

\noindent\textbf{Existing Work.} MMoE \citep{ma_modeling_2018}  is one of the most representative soft parameter sharing model structure in RSs which was proposed by Google in 2018. It is based on the Mixture-of-Experts (MoE) \citep{shazeer_outrageously_2017}, where the experts are shared across all tasks and a gating network is trained for each task. MMoE adapts the idea of integrated learning and uses several expert models to weight the influence on sub-tasks, with the gate mechanism trained to obtain influence factors of different experts. Finally, the weights of different experts are output by softmax. Since its introduction, several classic frameworks for MTRSs have been proposed based on the MMoE model. For example, in 2019, SNR \citep{ma_snr_2019} modularizes the shared low-level layers into parallel sub-networks and learns their connections for different optimization objectives. In 2020, PLE \citep{tang_progressive_2020} was proposed to solve the seesaw phenomenon where improving the performance of one task may hurt the performance of some other tasks. PLE separates task-common and task-specific parameters explicitly to avoid parameter conflict resulting from complex task correlations.

Recent advancements in MTRSs have yielded substantial progress, with many studies extending the foundational MMoE and PLE models. For example, Zhao et al. \citet{zhao_recommending_2019} integrate the wide\&deep model with MMoE, incorporating a shallow tower to mitigate selection bias. Ma et al. \citet{ma_online_2022} develop a large-scale online MTL framework for modeling multiple feed ads auction prediction tasks. CMoIE \citep{wang_multi-task_2022} refines MMoE model by introducing an expert communication module and conflict resolution module, then they can construct a set of high-quality and insightful expert sub-networks. Besides, to tackle the complexities of multi-scenario and multi-task challenges, Zou et al. \citet{zou_automatic_2022} propose an Automatic Expert Selection framework, which utilizes a sparse policy to adapt to task/scenario correlations. M\(^3\)oE\cite{zhang2024m3oe} utilizes three MoE modules to disentangle and capture common, domain-specific, and task-specific user preferences, thereby addressing complex interdependencies across multiple domains and tasks. Furthermore, several studies, such as \cite{gu_deep_2020, wu_multi-task_2022, wang_mtcut_2022, zhang_ctnocvr_2022} have made modifications to the basic MMoE and PLE architecture to suit specific tasks or scenarios. STAN\cite{li2023stan} identifies latent user lifecycle stages based on evolving user preferences for different tasks and uses these representations to enhance the performance of MTRSs. For embedding layer, STEM-Net\cite{su2024stem} introduces shared and task-specific embeddings to better capture task-specific user preferences and uses a customized gating network with stop-gradient operations to enhance embedding learning. HoME \cite{wang2025home} explores three anomalies, including expert collapse, expert degradation and expert underfitting, and evolves the traditional Mixture-of-Experts (MoE) paradigm into a hierarchy of multi-gate experts. Instead of utilizing a flat pool of experts, HoME organizes experts into multiple levels through cascaded gating mechanisms. This hierarchical structure enables high-level, task-specific experts to adaptively leverage knowledge from low-level, generic experts, thereby achieving more granular and efficient hierarchical knowledge transfer.

There has been a surge of interest in MTRSs that leverage user behavior sequence dependencies. MoSE \citep{qin_multitask_2020} incorporates long short-term memory to model sequential user behavior within the MMoE framework. AITM \citep{xi_modeling_2021} is specifically designed to capture long-path sequential dependencies among audience multi-step conversions for improving end-to-end conversion. These models represent important advances in the field and demonstrate the growing interest in developing more sophisticated techniques for MTRSs.


\noindent\textbf{Advantages of Expert Sharing.} 

\begin{itemize}
    \item Flexibility: Expert sharing provides a balance between task-specific representations (through task-specific experts) and shared representations (through task-shared experts).
    \item Efficiency: By routing tasks to specialized experts, the model can often achieve better performance with fewer resources compared to a soft parameter sharing model.
    \item Scalability: As new tasks are introduced, new experts can be added to the model without a complete redesign.
\end{itemize}

\noindent\textbf{Limitations of Expert Sharing.}

\begin{itemize}
    \item Model Complexity: Managing multiple experts and a gating mechanism can increase the complexity of the model, both in terms of architecture and training.
    \item Risk of Overfitting: If not enough data is available for each task, there's a risk of overfitting, especially in the task-specific experts.
    \item Gating Decisions: A poorly trained gating mechanism can lead to suboptimal expert selections, hampering performance.
\end{itemize}

\subsubsection{Sparse sharing}

Sparse sharing refers to an approach where only a subset of model parameters or layers are shared across tasks, while other parameters or layers are task-specific. It is an intermediate approach between fully shared (hard parameter sharing) and fully separate (soft parameter sharing) architectures.  The key characteristics of hierarchical sharing include the following:
\begin{itemize}
    \item Selective Sharing: Only certain parts of the model, which might be layers or parameters, are shared across tasks. This could be based on domain knowledge, heuristics, or learned through techniques like regularization.
    \item Task-specific Components: In addition to the shared parts, there are components of the model that are exclusive to individual tasks.
    \item Regularization or Constraints: Sparse sharing incorporates some form of regularization or constraints to determine which parameters or layers should be shared. This can be done using techniques like group lasso regularization, where groups of parameters are either entirely included or excluded from the shared representation. 
\end{itemize}


\noindent\textbf{Existing Work.} Due to the proportional growth of network parameters with respect to the total number of tasks, hard parameter sharing, soft parameter sharing, and expert sharing models suffer from computational and memory inefficiencies. Sparse sharing approaches \citep{sun_learning_2020} are proposed to mitigate the problem of parameter explosion. Instead of using an extra parameter space, sparse sharing approaches connect two sub-networks from the shared parameter space with independent parameter masks using binary variables. With the help of neural network pruning techniques, each task can extract the related knowledge for its own subnet and avoid the issue of excessive parameters. MSSM \citep{ding_mssm_2021} is a novel sparse sharing approach, which consists of two main components: a field-level sparse connection module (FSCM) and a cell-level sparse sharing module (CSSM). The FSCM automatically determines the importance of feature fields for each task using a sparse mask and allows for both task-specific and task-shared feature fields to be learned in an end-to-end manner. The CSSM implements an efficient sharing architecture by using a more fine-grained and cell-level sparse connection among sub-networks. GRec \cite{wang2023efficient} leverages a sparse mixture-of-experts (MoE) architecture and introduces a novel task-sentence routing strategy, enabling efficient cross-task generalization while maintaining high inference performance. MoME\cite{xu2024mome} leverages binary mask learning to extract expert sub-networks from an over-parameterized base network, effectively allowing for selective and sparse sharing of parameters. DSL \cite{wang2024dynamic} innovatively trains a lightweight sparse model from scratch, periodically evaluating and dynamically adjusting each weight’s significance and the model’s sparsity distribution during the training. Adap-SL \cite{zheng2025adaptive} also achieves sparse sharing by pruning, which proposes an adaptive structure learning approach to adaptively learn the optimal network structure, adjust the number of activated parameters

\noindent\textbf{Advantages of sparse sharing.} 

\begin{itemize}
    \item Flexibility: By allowing both shared and task-specific components, sparse sharing provides a flexible middle-ground between hard and soft parameter sharing, making it adaptive to the characteristics of the tasks at hand.
    \item Mitigates Negative Transfer: By not enforcing sharing of all components, sparse sharing can potentially reduce the risk of negative transfer, where the learning for one task adversely impacts another task.
    \item Efficient Resource Use: By sharing only the necessary components, the model can potentially use computational resources more efficiently.
\end{itemize}

\noindent\textbf{Limitations of Sparse Sharing.}

\begin{itemize}
    \item Determination of Shared Components: Deciding which components to share and which to keep task-specific can be challenging and may require domain knowledge, extensive experimentation, or sophisticated regularization techniques.
    \item Increased Complexity: Introducing mechanisms to decide which parts to share can add to the complexity of the model and training process.
\end{itemize}

\subsubsection{Hierarchical sharing}

Hierarchical sharing refers to a model structure where tasks are organized in a hierarchy based on their relationships or similarities. The key characteristics of hierarchical sharing include the following:
\begin{itemize}
    \item Task Relationships: Tasks are organized based on their inherent relationships
or similarities, which can be determined through domain knowledge or learned from data.
\item Shared Representations at Different Levels: Common features or parameters are shared at various levels of the hierarchy. Tasks at the lower levels of the hierarchy might share more parameters, while tasks at higher levels might have more distinct, task-specific parameters.
\item Layered Abstractions: Hierarchical sharing typically involves creating abstractions at different layers. Lower layers capture generic features that are commonly shared, while higher layers capture more task-specific features.
\end{itemize}


\noindent\textbf{Existing work.} MFH\citep{liu_multi-faceted_2022} is an efficient MTL model designed for large-scale tasks and it owns three major characteristics: Multi-Faceted, Hierarchical and Heterogeneous. MFH adopts a two-level tree architecture, which models the task relationships in both facets and shares the facet latent representations between tasks in a hierarchical fashion. This approach addresses the issue of local overfitting and data scarcity, particularly for tasks with few samples. Moreover, the MFH network is more heterogeneity-friendly and provides great flexibility for the model to better customize the tasks and generate further improvement. HMT-GRN \citep{lim_hierarchical_2022} learns the user-POI matrix and User-Region matrices of different levels simultaneously. When the model learns User-Region matrices, it conducts a hierarchical beam search (HBS) on the different region and POI distributions to hierarchically reduce the search space and predict the next POI. However, designing an effective and general hierarchical sharing approach is time-consuming. Recently, the concept of hierarchical sharing has been further extended to accommodate complex data structures and large-scale industrial architectures, moving beyond simple bottom-layer sharing. In the domain of graph neural networks, particularly for multi-task learning on heterogeneous graphs, Feng et al. \cite{feng2025hierarchical} introduced a hierarchical structure sharing mechanism to assist customer expansion tasks. By facilitating the sharing and interaction of structural information across different levels of abstraction within the heterogeneous graph, this method enhances the model's capacity to capture complex cross-task relationships. Concurrently, to address the intricate task dependencies inherent in large-scale industrial recommendation scenarios (e.g., Kuaishou), HoME \cite{wang2025home} evolves the traditional Mixture-of-Experts (MoE) paradigm into a hierarchy of multi-gate experts. Instead of utilizing a flat pool of experts, HoME organizes experts into multiple levels through cascaded gating mechanisms. This hierarchical structure enables high-level, task-specific experts to adaptively leverage knowledge from low-level, generic experts, thereby achieving more granular and efficient hierarchical knowledge transfer.

\noindent\textbf{Advantages of Hierarchical Sharing.} 

\begin{itemize}
    \item Structured Sharing: By organizing tasks hierarchically, the model can share parameters in a more structured manner, aligning closely with the relatedness of tasks.
    \item Mitigates Negative Transfer: By sharing parameters at different levels of the hierarchy, hierarchical sharing can reduce the risk of negative transfer between unrelated tasks.
    \item Scalability: Hierarchical sharing can be more scalable to a large number of tasks, as adding new tasks might simply involve placing them in the existing hierarchy.
    \item Improved Generalization: Hierarchical structures can leverage the shared information in upper layers for better generalization on tasks, especially when there's limited data for some tasks.
\end{itemize}

\noindent\textbf{Limitations of Hierarchical Sharing.}

\begin{itemize}
    \item Complexity: Building and training a hierarchical model can be more complex than flat multi-task models.
    \item Determining the Hierarchy: In some domains, establishing a clear hierarchy of tasks might be challenging, and it might require domain expertise.
    \item Flexibility: While the hierarchical structure provides a structured way of sharing, it might not always be the most flexible, especially if the relationships among tasks change or if new tasks are introduced that don't fit neatly into the existing hierarchy.
\end{itemize}
\subsubsection{Dynamic Parameter Sharing}

Dynamic parameter sharing refers to a method in which the sharing of parameters across tasks is adaptively adjusted during training to meet the specific needs of each task. Unlike static parameter sharing, which relies on predetermined and fixed parameter allocations, dynamic parameter sharing allows the model to continuously learn and refine the optimal way to share information across tasks in real time, enhancing flexibility and adaptability. The key characteristics of dynamic parameter sharing include the following:

\begin{itemize} 
    \item \textbf{Adaptive Sharing Mechanism}: Parameters are shared in an adaptive manner, with the sharing decisions being adjusted during training. This ensures that the model continuously refines its parameter-sharing strategy to maximize performance. 
    \item \textbf{Task-Specific Customization}: Each task can learn to dynamically select shared parameters that are most relevant, allowing for more customized learning. For instance, attention-based mechanisms or gating networks can be used to enable each task to selectively share parameters.
\end{itemize}

\noindent\textbf{Existing Work.} SNR \cite{ma_snr_2019} introduces flexible parameter sharing through modular sub-networks and learnable binary encoding variables, enhancing both the trainability and performance of MTL models. Similarly, Jiang et al. \cite{jiang2024automatic} present an innovative automated MTL framework, AutoMTL, which leverages neural architecture search (NAS) to design optimal expert architectures and sharing strategies. They further propose an efficient search algorithm that explores the vast search space while progressively discretizing it to improve both the efficiency and accuracy of the search process. Adap-SL \cite{zheng2025adaptive} designs an adaptive structure learning approach to adaptively learn the optimal network structure, adjust the number of activated parameters, and determine which knowledge needs to be transferred between the prediction model and imputation model.

\noindent\textbf{Advantages of Dynamic Parameter Sharing.} 

\begin{itemize} 
    \item \textbf{Task-Specific Optimization}: Dynamic parameter sharing allows the model to adaptively determine the level of sharing for each task, leading to better optimization for tasks with different characteristics. 
    \item \textbf{Reduced Negative Transfer}: By dynamically adjusting shared parameters, the model reduces the risk of negative transfer, as it only shares parameters that are beneficial for a particular task. 
    \item \textbf{Flexibility}: This approach provides greater flexibility compared to static sharing methods, making it suitable for scenarios where task relationships change or when new tasks are introduced. 
\end{itemize}

\noindent\textbf{Limitations of Dynamic Parameter Sharing.} Although \cite{jiang2024automatic} optimizes search efficiency, neural architecture search (NAS) itself still requires a lot of computing resources, especially when facing complex tasks and large-scale data, the cost is still high.

\subsection{Transfer Learning Architecture}

Multitask learning aims to exploit the similarities and dependencies among a set of related tasks to enhance the performance of each task. Information transfer can occur at different levels, including parameter sharing, intermediate representation sharing, or certain components sharing of the model. It is worth noting that we define intermediate representation sharing and certain components sharing  as transfer learning architectures within MTRSs. For example, MRAN \citep{zhao_multiple_2019} consists of three relationship learning modules. In the task-task transfer relationship learning module, it exploits the self-attention mechanisms to control the positive and negative knowledge transfer among tasks. AITM \citep{xi_modeling_2021} proposes an adaptive information transfer multi-task framework, different conversion stages of different audiences need to transfer different information from the former step to the latter step. Based on the expert sharing model, MNCM \citep{wu_mncm_2022} introduces Task-level Information Transfer Module (TITM) and Expert-level Information Transfer Module (EITM) to adaptively control the amount of information that is transferred to the next task. Inspired by knowledge distillation, CrossDistil\citep{yang_cross-task_2022} facilitates the transfer of classification knowledge into the top layers, and they adopted calibrated knowledge distillation to transfer knowledge from augmented tasks (teacher) to original tasks (student).  SeqMT\citep{pentina2015curriculum}, MTSIT\citep{xu_multi-task_2018} processes multiple tasks sequentially to transfer information between sequential tasks. Notably, different from the existing MTL methods, they aims to learn a robust classification model by exploiting both the complexity of instances and that of tasks.

\subsection{Adversarial Learning Architecture}

Adversarial learning networks have been widely used in image recognition, style migration, speech recognition and etc. Recently, adversarial learning-based RSs have gained increasing attention in the research community. In 2017, Wang et al. \citet{wang_irgan_2017} proposed IRGAN which introduced GANs to information retrieval (IR) for the first time. This paper showed that IRGAN can be applied to web search, item recommendation, and question answering. In GAN-based approaches, the discriminator network provides feedback on the relevance of the recommended items to the user, which helps improve the accuracy and relevance of the recommendations. In addition to IRGAN, some adversarial learning-based RSs have been proposed, such as Raj et al. \citet{raj2023multi} develop an Adversarial Shared Private Multi-Task model for solving multi-modal movie recommendation. This model utilize adversarial training to learn shared representations across tasks while also preserving privacy by keeping task-specific information separate. The adversarial training is used to encourage the shared feature extractor to extract features that are useful for rating prediction task and genre prediction task while also discouraging it from extracting task-specific information. Su et al. \citet{su2022cross} propose an adversarial adaptation framework which solve domain shift problem and mitigate the imbalanced target distribution. In this framework, it deploys a domain discriminator to specifically align target domain with source domain and model CTR task and CVR task of the source model to improve the CVR task on target domain. AFT \cite{hao2021adversarial} also employs generative adversarial networks for feature translation between domains. Overall, adversarial learning has shown promise in cross-domain scenarios. Su et al. \cite{su2023task} construct an adversarial training paradigm with all task feature discriminators and the feature embedding network. Through a two-player min-max game between the feature discriminator and the feature embedding network, the feature embedding network is guided to encode domain-shared joint features for each task. AECM\cite{zhang2024adversarial} introduces a novel dual adversarial module, operating on both the space and task level. It can denoise the click propensity and capture counterfactual embeddings to optimize the CVR task. Multi-ACG\cite{ma2025multitask} integrates self-supervised and adversarial learning to generate "hard samples" for robust representation learning in graph-based recommendation. Other approaches leverage adversarial training to enhance robustness; for instance, CLAM\cite{xiang2025contrastive} employs an adversarial occlusion module to adaptively generate challenging embedding masks during contrastive learning, ensuring the primary recommendation model learns representations resilient to semantic perturbations.

\subsection{Large Language Models-based Architecture}

The successful application of pre-trained large language models(LLM) such as BERT, GPT, and T5 in the field of natural language processing (NLP) has exhibited their powerful capabilities in semantic understanding and generative modeling. The integration of Large Language Models (LLMs) into RSs has revolutionized the architecture of MTRSs \cite{lin2025can, zhao2024recommender}. Depending on the role LLMs play in the inference process, we categorize existing architectures into two paradigms: \textit{LLM as Recommender} and \textit{LLM as Enhancer}. The former unifies multiple recommendation tasks into a single generative framework, fundamentally reshaping the MTL paradigm. The latter integrates LLMs as auxiliary modules into traditional recommendation models to enhance representation learning or data quality.

\subsubsection{LLMs as Recommender}

In this paradigm, LLMs act as the central decision-maker, unifying diverse recommendation tasks into a single sequence-to-sequence format. Representative encoder–decoder models such as M6-Rec \cite{cui2022m6}, P5 \cite{geng2022recommendation}, InstructRec \cite{zhang2025recommendation}, and URM \cite{jiang2025large} leverage instruction tuning and task-specific prompting templates to align LLMs with various recommendation tasks. Recent works \cite{luo2025recranker, meng2025recrankereval} further explore LLMs' potential for top-N recommendations by jointly modeling pointwise scoring, pairwise comparison, and listwise ranking within a unified prompting framework. CKF \cite{zhao2025collaborative} incorporates collaborative embeddings into prompt templates via a personalized meta-mapping network and employs Multi-LoRA to separately model shared and task-specific components, achieving improved multi-task performance. UniMP \cite{wei2024towards} establishes a unified paradigm for multi-modal personalization systems based CLIP ViT-L model, which can handle a wide range of recommendation tasks, including item recommendation, product search, preference prediction, explanation generation, and user-guided image generation. Deng et al. \cite{deng_unified_2022} developed a multi-goal conversational recommender system, unifying four conversational sub-tasks within a consistent Seq2Seq framework using prompt-based learning strategies.

Furthermore, LLMRec \cite{lyu2023llm} evaluates popular LLMs (e.g., ChatGPT, LLaMA, ChatGLM) across multiple recommendation tasks, revealing their strengths in explainability-oriented tasks and limitations in accuracy-driven tasks. These findings offer valuable insights into when and how LLM-only architectures are most suitable for MTRSs.

\subsubsection{LLM as Enhancer}

We analyze the role of LLMs in MTRSs from three complementary perspectives: LLMs as Feature Encoders, LLM-driven Data Augmentation, and LLM-based Explanation. Together, these perspectives provide a systematic understanding of how LLMs enhance representation learning, enrich supervision signals, and improve interpretability in MTRSs.

Recent studies increasingly adopt LLMs as semantic feature encoders to enrich representation learning in MTRSs. These methods leverage LLMs to extract high-level semantic embeddings from user profiles, items, or contextual descriptions, which are then fused with collaborative-filtering representations. For example, LLM4MSR \cite{wang2024llm4msr} uses LLMs to distill scenario-aware and user-specific knowledge and integrates it into MSR models via hierarchical meta-networks, achieving superior performance. Gone et al.\cite{gong2023unified} proposes S\&R Multi-Domain Foundation, a framework that uses LLMs to extract domain-invariant textual features and fuses them with ID-based and task-specific signals for unified multi-domain search and recommendation. LEMON \cite{wang2025enhanced} designs a hierarchical LLM-driven emotion extraction pipeline to generate semantically rich song embeddings, and then predicts multiple task by the  multifaceted embedding.

A growing body of research explores the use of LLMs to generate synthetic or enriched interaction data to address challenges such as data sparsity, long-tail distributions, and cold-start problems in MTRSs. For instance, LLM4DSR \cite{wang2025llm4dsr} utilizes LLMs to identify and filter noisy behaviors within sequences, or to rewrite semantic labels for unreliable actions. SampleLLM \cite{gao2025samplellm} applies LLM-driven synthesis and augmentation to tabular data (user features, item features, and labels), which is particularly useful for expanding training datasets in sparse or cold-start scenarios. The generated datasets can then be used for MTRSs. In cases where target domain data is insufficient, LEKA \cite{zhang2025leka} leverages LLMs to retrieve and harmonize relevant data from external knowledge bases or open networks, effectively addressing domain shift issues in cross-domain and new task settings. Additionally, LLMRec \cite{wei2024llmrec} employs three simple yet effective LLM-based graph augmentation strategies to enhance recommendation performance, while also tackling the challenges posed by sparse implicit feedback and low-quality side information. MADRec \cite{park2025madrec} generates structured profiles via aspect-category-based summarization and applies RE-RANKING to construct high-density inputs for the direct recommendation, sequential recommendation and explanation generation tasks. In summary, LLM-driven data augmentation has become a powerful tool for mitigating data sparsity of MTRSs.

LLMs have been widely used to improve the explainability of recommendation systems, often by generating natural language rationales or performing reasoning over knowledge graphs. These explanation signals can act as auxiliary tasks in MTL. Representative works include LLM-PKG \cite{wei2024llmrec}, which integrates LLM-based knowledge-graph reasoning with preference explanation; LANE \cite{zhao2024lane} generates high-quality explanations for ranking tasks by utilizing structured reasoning chains. These approaches demonstrate that incorporating LLM-based explanations can improve transparency while providing complementary semantic signals that benefit MTRSs.

\subsection{Summary}

\begin{table*}[t]
\centering
\small
\caption{Comparison of Four Architectural Paradigms in Multi-Task Recommendation Systems}

\begin{tabular}{p{2.5cm} p{4.6cm} p{4.8cm} p{3.8cm}}
\toprule
\textbf{Architecture} &
\textbf{Strengths} &
\textbf{Weaknesses} &
\textbf{Applicable Scenarios}
\\ \midrule

\textbf{Parameter Sharing} &
Simple and efficient.\newline
Good sample efficiency.\newline
Strong shared semantics (user/item).\newline
Works well for related tasks.
&
Risk of negative transfer. \newline
Task interference.\newline
Dominated by high-frequency tasks.\newline
Weak for heterogeneous behaviors.
&
Multi-behavior RSs;\newline
multi-objective CTR-CVR;\newline
industrial MTL.
\\

\textbf{Transfer Learning} &
Enhances sparse tasks.\newline
Supports asymmetric transfer.\newline
Improves target-task performance.\newline
Clear task dependency modeling.
&
Sensitive to task order.\newline
Potential negative transfer.\newline
Requires careful module design.
&
Sequential recommedantion;\newline
cross-domain RS;\newline
multi-stage task pipelines.
\\

\textbf{Adversarial Learning} &
Improves robustness.\newline
Reduces bias and mismatch.\newline
Effective under domain shift.\newline
Strong generalization.
&
Training instability.\newline
Optimization difficulty.\newline
May suppress task-specific signals.\newline
Higher computation cost.
&
Cross-domain RS;\newline
debiasing;\newline 
privacy-preserving RS.

\\

\textbf{LLM-based} &
Strong generalization ability.\newline
Explains user intent.\newline
Works in sparse scenarios.\newline
Effective for long-tail tasks.
&
Very high computational cost.\newline
Hard to align with collaborative filters.\newline
Hallucination risk.\newline
Limited ranking accuracy.
&
text-based RSs;\newline
cross-domain RSs;\newline
cold-start RSs;\newline
multimodal recommendation.
\\
\bottomrule
\end{tabular}
\end{table*}

Most existing MTRSs focus on parameter sharing and transfer learning structures, while adversarial learning structures have received comparatively less attention. This may be attributed to the risk of task interference in adversarial MTL, particularly when tasks exhibit significant differences, potentially reducing overall task performance. To address these limitations and explore future directions, we identify key areas of research for various MTRSs:

\paragraph{1) Parameter Sharing-Based MTRSs}

Parameter sharing remains a widely used paradigm in MTRSs due to its ability to leverage shared knowledge across tasks. However, several challenges and opportunities warrant further exploration:

\begin{itemize}
    \item \textbf{Dynamic Parameter Sharing}: Investigating mechanisms to dynamically adjust the scope and method of parameter sharing based on the characteristics and requirements of individual tasks. Such mechanisms can enhance model adaptability and efficiency.
    \item \textbf{Efficient Sharing Structures}: Designing more efficient parameter sharing architectures to improve knowledge transfer and generalization across diverse tasks.
    \item \textbf{Cross-Domain Applications}: Extending parameter sharing techniques to cross-domain recommendation scenarios to address data sparsity and enhance generalization across domains.
\end{itemize}

\paragraph{2) Transfer Learning-Based MTRSs}

Transfer learning offers a robust framework for transferring knowledge between tasks, but its effectiveness depends on mitigating challenges like negative transfer. Key research directions include:

\begin{itemize}
    \item \textbf{Optimizing Knowledge Transfer}: Developing efficient mechanisms to transfer knowledge between tasks by quantifying and leveraging task relationships, as discussed in Section \ref{sec3} on task relationship discovery.
    \item \textbf{Mitigating Negative Transfer}: Identifying the causes of negative transfer and designing strategies to minimize its impact, ensuring that knowledge transfer benefits all tasks.
\end{itemize}

\paragraph{3) Adversarial Learning-Based MTRSs}

Adversarial learning has been primarily utilized in scenarios such as detecting fraudulent users or malicious behavior in RSs. Future research should focus on:

\begin{itemize}
    \item \textbf{Personalized Recommendations within Adversarial Frameworks}: Investigating methods to ensure effective personalized recommendations in adversarial learning settings, addressing challenges like task interference.
    \item \textbf{User Privacy and Data Security}: Exploring techniques to integrate adversarial learning with robust privacy-preserving mechanisms, ensuring secure and ethical recommendation practices.
\end{itemize}

\paragraph{4) LLM-Based MTRSs}

As pre-training technologies evolve, large language models (LLMs) are becoming increasingly central to personalized recommendation systems. LLM-based MTRSs exhibit several advantages and open avenues for further exploration:

\begin{itemize}
    \item \textbf{Enhanced Generalization}: By leveraging LLMs' capacity to understand and generate natural language, MTRSs can significantly improve performance in sparse domains. For instance, when user interaction data in a specific domain is limited, the model can utilize knowledge from other tasks or domains to fill these gaps, resulting in more accurate recommendations.
    \item \textbf{Advanced Personalization}: LLMs enable deeper understanding of diverse user needs, enriching user interactions and enhancing recommendation precision.
    \item \textbf{Future Prospects}: LLM-powered RSs are expected to pioneer intelligent systems capable of adapting to dynamic user needs and offering more personalized, context-aware recommendations. These advancements will likely improve user satisfaction and open new frontiers in personalized recommendation.
\end{itemize}

By addressing these research opportunities, the field of MTRSs can progress toward more adaptable, secure, and intelligent systems, leveraging advancements across parameter sharing, transfer learning, adversarial learning, and LLM-based approaches to enhance recommendation quality and user satisfaction.

\section{Optimization Strategy of MTRSs}\label{sec5}

Multi-task recommendation requires the simultaneous optimization of multiple tasks, and joint optimization of these tasks poses several challenges: (i) the weight allocation for different tasks when converting MTL into single-objective optimization, (ii) the conflict across the gradient of multiple tasks, and (iii) the trade-off between multi-objectives. 

\subsection{Loss weight}

The loss weights serve like a tuning knobs to tune the importance of a specific task among all other tasks. Increasing the weight of one task usually leads to a bit better performance of this task in the cost of other tasks’ performance. Regarding how to balance the loss weight, a common solution is to combine multi task losses and perform a weighted sum of the losses of different tasks. However, most exiting loss weights solution can not be suitable for online advertising. Ma et al. \citet{ma_online_2022} define a single MTL objective which can guide the task loss weight tuning. Li et al. \citet{li_spex_2022} design an automatic task balancing mechanism to weigh task losses via capturing homoscedastic uncertainty in a MTL setting.

\subsection{Gradient}

In some scenarios, MTL may not achieve satisfactory performance due to the presence of gradient conflicts among different tasks. This means that the gradient directions of different tasks form an angle larger than 90 degrees. To address this issue, Yu et al. \citet{yu2020gradient}  propose PCGrad(projecting conflicting gradients) to project the gradient of a task onto the normal plane of another conflicting task's gradient. This gradient correction algorithm avoid gradient conflicts by constructing new gradient update directions, but it is easy to deviate from the goal of minimizing average loss, making it difficult for the final result to converge to the global optimal. To solve this problem, Liu et al. \citet{liu_conflict-averse_2021} introduce CAGrad that seek the gradient direction within neighborhood of the average gradient and maximize the worst local improvement of any objective. Later, Bi et al. \citet{bi_mtrec_2022} merge the gradients of auxiliary tasks and applied a scaling factor to adaptively adjust their impact on the main tasks, followed by applying gradient sharing between the main tasks and the merged auxiliary task. 

Furthermore, the gradient magnitudes for different tasks may vary widely, which can negatively affect the performance of the model. To tackle this issue, GradNorm \citep{chen_gradnorm_2018} presents a gradient normalization algorithm that dynamically balances training in deep multi-task models by tuning gradient magnitudes. Liu et al. \citet{liu_federated_2022} propose a communication-efficient federated optimization method, named MTgTop-k S-SGD, which applies different update strategies on two parts of the parameter. Specifically, the local SGD updates the task-specific parameters, while a tree-based distributed gradients aggregation scheme is applied to update the shared parameters. He et al. \citet{he_metabalance_2022} design a method called MetaBalance, which introduces a relaxation factor to flexibly control the gradient magnitude proximity between auxiliary and target tasks. This method dynamically balances the auxiliary gradients throughout the training process and adaptively adjusts the gradients for each part of the network.

\subsection{Multi-objective Optimization}

MTL optimization can also be regarded as a multi-objective optimization problem in essence, because there may be conflicts between different tasks, so it is necessary to make trade-offs between multiple tasks and find a Pareto optimal solution. In \cite{sener_multi-task_2018}, they transform MTL problems into multi-objective optimization with constraints. Then they introduce multiple gradient descent algorithm (MGDA) to reduce training cost and apply Frank-Wolfe-based optimizer to yield a Pareto optimal solution under realistic assumptions. However, the solutions obtained by existing methods are sparse and discrete. To fill the gap, Ma et al. \citet{ma_efficient_2020} propose a novel approach for generating locally continuous Pareto sets and Pareto fronts when transforming MTL problems into multi-objective optimization problems. This opens up the possibility of continuous analysis of Pareto optimal solutions in machine learning problems.

\section{Training Paradigms}

In MTRSs, effective training strategies are crucial for jointly optimizing multiple objectives across tasks and scenarios. This section introduces three key training strategies: joint learning, reinforcement learning, and pre-training with fine-tuning.

\subsection{Joint Learning}

Joint learning is one of the most widely adopted training paradigms in MTRSs. Its core goal is to improve the overall performance by optimizing multiple tasks simultaneously within a unified framework and exploiting the potential correlations between tasks. It is particularly effective for tasks with clear relationships and abundant data but faces challenges such as gradient conflicts and task imbalance.

Most MTRSs adopt a joint learning training approach, where the losses of individual tasks are weighted and summed to form a joint loss function. This training strategy enables knowledge sharing and regularization among tasks by optimizing multiple task objectives simultaneously within a shared parameter framework. Representative methods, such as those proposed in \cite{chen_multi_2021, chen_co-attentive_2019, wang_escm2_2022, li2023stan}, leverage this strategy to enhance overall model performance. While this method is simple and efficient, it may face challenges such as task conflicts or negative transfer due to improper weight allocation among tasks.

\subsection{Reinforcement Learning}

Reinforcement Learning (RL) has shown great potential in MTRSs by modeling sequential user behaviors as a Markov Decision Process (MDP) and leveraging policy networks to optimize task interactions. In multi-task settings, RL is primarily utilized to dynamically balance task objectives, optimize long-term rewards, and improve overall system performance.

For example, Liu et al. \citet{liu_multi-task_2023} proposed RMTL, an RL-enhanced multi-task recommendation framework that formulates multi-task prediction data as an MDP. The framework dynamically updates task weights in the overall loss function, allowing adaptive optimization of multiple task objectives. Similarly, Zhang et al. \citet{zhang_multi-task_2022} introduced BatchRL-MTF, a Batch RL-based Multi-Task Fusion framework, which formulates the task fusion process within a recommendation session as an MDP. This method focuses on optimizing long-term user satisfaction across multiple tasks by leveraging batch reinforcement learning techniques.

To address the issue of policy-environment mismatch caused by the lack of real-time interaction during offline training in large-scale RSs, Liu et al. \cite{liu2024off} proposed a progressive training mode. This strategy divides the training process into multiple iterations of online data exploration and offline model training. An efficient exploration policy is employed to guide the model toward the optimal policy, enabling more effective task collaboration and convergence.

While the integration of RL into multi-task training in RSs is still in its early stages, these works illustrate the potential of RL to address key challenges, such as dynamic task balancing, long-term reward optimization, and policy convergence. As RL has been successfully applied to broader MTL scenarios, its extension to RSs holds significant promise for improving multi-task training strategies.

\subsection{Pre-training and Tune}

Pre-training and fine-tuning have emerged as a powerful paradigm in natural language processing, enabling models to learn generalizable representations from large-scale data and adapt them effectively to downstream tasks. This paradigm enables models to learn generalizable representations from large-scale data during the pretraining phase and adapt effectively to downstream tasks through fine-tuning. Inspired by this success, similar techniques have been adopted in RSs to address challenges such as data sparsity, task heterogeneity, and the need for efficient domain adaptation. By leveraging pre-trained models, RSs can capture universal patterns in user behavior and item characteristics, enabling more accurate and personalized recommendations across a wide range of tasks and domains.

Several representative frameworks demonstrate the effectiveness of this paradigm. For instance, M6-Rec \cite{cui2022m6} introduced option tuning, an advanced prompt-tuning method that significantly reduces training costs for downstream tasks while minimizing inference time and model size through techniques such as pruning and parameter sharing. Similarly, P5 \cite{geng2022recommendation} adopts a unified training strategy that transforms recommendation tasks into a text-to-text paradigm. During pretraining, user-item interactions and metadata are converted into natural language sequences, and the model is trained with a language modeling objective to capture deep semantic representations and facilitate knowledge sharing across tasks. By using personalized prompts during fine-tuning, P5 adapts effectively to specific recommendation tasks, supporting minimal fine-tuning while achieving efficient task adaptation, strong generalization, and seamless integration with multimodal data. To address task irrelevance and enhance training efficiency, MPT-Rec \cite{bai2025efficient} introduces a novel two-stage prompt-tuning MTL framework. This framework employs task-aware prompts to effectively transfer knowledge from older tasks to new ones, improving the overall training process. Additionally, CSMF \cite{deng2025csmf} proposes a method that enhances both retrieval efficiency and serving performance in multi-objective retrieval scenarios. The CSMF approach incorporates an iterative training process that cycles through the stages of pre-training, selective masking, accuracy recovery, and fine-tuning to optimize model performance.

Other notable frameworks include UPRTH \cite{yang2024unified}, which proposes a model-agnostic multi-task pretraining framework. This approach generalizes all pretext tasks as hyperedge predictions within task hypergraphs, leveraging hypergraph structures to encode high-order relationships across tasks for robust MTL and improved knowledge sharing. MultiGPrompt \cite{yu2024multigprompt} addresses task interference and effective knowledge transfer for few-shot learning in a graph-based multi-task pretraining framework. It introduces pretext tokens to synergize multiple pretext tasks during pretraining, reducing interference and enabling collaboration. For downstream adaptation, it employs a dual-prompt mechanism, consisting of a composed prompt to extract task-specific knowledge and an open prompt to capture global inter-task knowledge. 

These frameworks highlight the effectiveness of pre-training and fine-tuning as a training strategy for MTRSs. They unify diverse recommendation tasks within a single architecture, improve generalization across tasks, and provide an efficient mechanism to integrate multimodal data. This paradigm not only enhances task performance but also establishes a scalable foundation for future developments in MTRSs.

\section{Synergy with Other Techniques} \label{sec6}

With the growing complexity of real-world problems, it is increasingly common to encounter situations where a single learning paradigm is not sufficient to capture all the relevant aspects of the problem. In such cases, the combination of multiple learning paradigms can be a promising approach to achieve better performance. We explore the combination of MTL with different learning paradigms including knowledge graphs (KGs), graph neural networks (GNNs), reinforcement learning (RL) and multi-view learning. These learning paradigms have wide applications in various fields and offer unique methods to solve recommendation problems. By combining MTL with these learning paradigms, we can better utilize the strengths of each paradigm and achieve more efficient and accurate learning. This chapter will discuss how this combination can improve model performance  and investigate how this can lead to improved performance and more efficient use of resources.

\subsection{Knowledge Graphs}

RSs frequently encounter issues with data sparsity and cold start problems. While MTL can alleviate data sparsity to some extent, knowledge graphs are typically employed to mitigate cold start issues by providing prior knowledge. Therefore, numerous studies have combined MTL with KGs to construct recommendation models. For example, Zhang et al. \citet{zhang_knowledge-enhanced_2022} utilize medical knowledge graphs and medicine attribute graphs to learn embedding representations in a multi-task framework. DFM-GCN \citep{xiao_dfm-gcn_2022} applies DeepFM to extract high-level information and uses graph convolutional networks (GCN) to encode item embeddings within a knowledge graph. FairSR \citep{li_fairsr_2022} encodes user and item attributes as well as their relationships into entity embeddings in a multi-task end-to-end model. KGMTL4Rec \citep{dadoun_predicting_2021} leverages a MTL model to learn vector representations of KG entities containing travel-related information and then computes travel destination recommendation scores between travelers and destinations. Additionally, several studies \cite{wang_enhanced_2021,gao_enhanced_2023} utilize KGs as a source of side information and incorporate KG embedding tasks to assist recommendation tasks.

\subsection{Graph Neural Networks}

Graph neural networks (GNNs) have become increasingly popular in RSs to effectively model interactions between users and items. To address the challenge of data sparsity, Lim et al. \citep{lim_hierarchical_2022} introduce the Hierarchical Multi-Task Graph Recurrent Network (HMT-GRN), which leverages MTL to construct distinct User-Region matrices characterized by lower sparsity levels. Jian et al. \citep{jian_multi-task_2022} proposed the M-HetSage model, a unified multi-task heterogeneous graph neural network designed to capture complementary information from diverse customer behavior data sources. More recently, graph convolutional neural networks (GCNs) have also been extensively applied in MTRSs. For example, Wang et al. \citep{wang_entity-enhanced_2022} explore the effectiveness of GCNs alongside MTL techniques in capturing the complex tripartite relations among users, items, and entities. Yin et al. \citep{yin_recommendation_2022} incorporate knowledge graph embeddings as auxiliary tasks, using GCN to enhance the recommendation performance. In order to generate explanations alongside recommendations, ExpGCN \citep{wei_expgcn_2022} aims to create distinct node representations for both explanation ranking and item recommendation tasks by aggregating information from separate subgraphs. Furthermore, Yu et al. \cite{yu2024multigprompt} studied MultiGPrompt, an approach for multi-task pre-training on graphs to effectively transfer multi-task knowledge to downstream few-shot tasks. Feng et al. \cite{feng2025hierarchical} introduced a hierarchical structure sharing mechanism to assist customer expansion tasks. By facilitating the sharing and interaction of structural information across different levels of abstraction within the heterogeneous graph, this method enhances the model's capacity to capture complex cross-task relationships. Collectively, these works illustrate the significant progress in leveraging GNNs for MTRSs and enhancing model performance by incorporating knowledge from various related tasks.

\subsection{Reinforcement Learning }

Reinforcement learning has been applied in recommendation systems to obtain the optimal recommendation policy for long-term user satisfaction through interactive rewards. For example, Chen et al. \citet{chen_knowledge-based_2021} apply the actor-critic framework of RL to learn a dialogue policy in conversation scenarios of MTRSs.  However, applying RL in the large-scale online recommendation systems to optimize long-term user satisfaction remains a nontrivial task. Zhang et al. \citet{zhang_multi-task_2022} mainly focus on multi-task fusion algorithms and formulates the multi-task fusion task as a Markov Decision Process (MDP) within a recommendation session. They propose a Batch RL-based Multi-Task Fusion framework (named BatchRL-MTF) to optimize long-term user satisfaction. Moreover, RMTL \cite{liu_multi-task_2023} leverages RL in the optimization of MTL and improves the prediction performance of multi-tasks by generating adaptively adjusted weights for the loss function in a reinforcement learning manner.

\subsection{Multi-view Learning}

There are a wide variety of applications of multi-view MTL \citep{li_interactions_2018,lu_multilinear_2017}. Jin et al. \citet{jin_multi-task_2014} propose a Multi-tAsk MUlti-view Discriminant Analysis (MAMUDA) method to deal with the scenarios where the tasks with several views correspond to different set of class labels. Wang et al. \citet{wang_m2grl_2020} propose a multi-task multi-view graph representation learning framework (M2GRL) to learn node representations from multi-view graphs for web-scale RSs. Yin et al. \citet{yin_learning_2020} design a new transformer-based hierarchical encoder to model different kinds of sequential behaviors, which can be seen as multiple distinct views of the user’s searching history. And this method proposes a multi-view multi-task attentive framework to learn personalized query auto-completion models. 

\subsection{Summary}

The integration of MTL with diverse learning paradigms has unlocked novel prospects and dimensions for enhancing the performance of recommendation systems. Such synergies enable a more nuanced understanding of the interconnections between different tasks, uncovering complementary information that bolsters the overall quality of recommendations. For instance, knowledge graphs, through entity linking and relational reasoning, augment user profiles and item descriptions, thus facilitating the recommendation system's ability to discern cross-domain user interests and tailor personalized suggestions. Graph neural networks, leveraging graph embeddings, capture the intricacies of user-item interactions and the fine-gained relationship of user-user and item-item, further refining feature representation. The convergence of MTL with reinforcement learning optimizes static data and adjusts dynamically to changes in user preferences. The integration of reinforcement learning also strategically focuses on immediate user satisfaction and nurtures long-term user engagement. Multi-view learning is particularly adept at addressing multi-scenario recommendations, where it provides customized solutions across diverse user engagement platforms. 

As technological advancements continue, it becomes imperative to explore advanced learning paradigms to tackle complex challenges in MTRSs. For example, large language models could be harnessed to delve into comprehensive user preferences and profiles. Geng et al. \citet{geng2022recommendation} propose a novel recommender system modeling paradigm, which model various recommendation tasks in a unified pretrain, prompt, and predict conditional language generation framework. They transform the recommendation task into a natural language sequence through the designed personalized prompts template. Furthermore, meta-learning and few-shot learning approaches could provide innovative pathways for enhancing the efficiency and effectiveness of recommendation systems in data-sparse environments.

\section{Application Scenario}\label{sec7}

In recent years, MTRSs obtain significant attention due to its potential to address various challenges in real-world recommendation scenarios. In this section, we will discuss the application of MTRSs and summarize some common task combinations that appear in different recommendation scenarios, including but not limited to E-commerce recommendation, advertising recommendation, explainable recommendation, conversational recommendation and other recommendation. This information provides researchers with some prior works about task combinations in MTRSs for reference, and subsequent researchers can further investigate more task combinations in different scenarios.

 

\subsection{E-commerce Recommendation}

E-commerce recommendation is a prevalent recommendation scenario that aims to recommend commodities, services, or content that users may be interested in. It is widely applied in e-commerce, entertainment, and other fields. In the existing research on E-commerce recommendation, many predictive tasks have been combined to improve the recommendation performance, as shown in Table \ref{tab1}. Among these tasks, CTR and CVR are the most common optimization goal. CTR refers to the ratio of users clicking on an item, which can reflect whether the item recommended by the recommender system is attractive to the user. CVR refers to the proportion of users who finally purchase after clicking on an item, which can reflect whether the items recommended by the recommendation system meet the needs of users. Ma et al. \citet{ma_modeling_2018} first build a single multi-gate mixture-of-experts(MMoE) model that learns multiple goals and tasks simultaneously. Then, Tang et al. \citet{tang_progressive_2020} design task-shared experts and task-specific experts based on MMoE model with the aim of improving utilization of expert modules and reducing negative transfer. Later, Xiao et al. \citet{xiao_ncs4cvr_2023} propose a neuron-connection level sharing mechanism which can automatically decide which neuron connection to be shared without artificial setting. Besides, click-through\& conversion rate(CTCVR) is introduced as an auxiliary task to train with CTR task, thus the derived task CVR can be estimated over the entire space, which can eliminate SSB problem. In ESM\(^2\) \citep{wen_entire_2020}, click-through\&action conversion rate(CTAVR) is also added to assist the main task. In addition to the above tasks, there are some common user feedback that are regarded as tasks in e-commerce, such as buy, like, save, add-to-cart(ATC) and Deterministic Micro set(D-Mi, including clicking item’s pictures, checking the Q\&A details of an item, chatting with sellers, reading an item’s comments, clicking an item’s carting control button). Ni et al. \citet{ni_perceive_2018} propose more prediction tasks in E-commerce recommendation to improve the performance, including Learning to Rank(L2R), Price Preference Prediction(PPP), Fashion Icon Following Prediction(FIFP) and Shop Preference Prediction (SPP). Through modeling these tasks, the user's preference representation can be better learned. In order to solve the the delayed feedback problem of CVR task, non-delayed positive rate(NDPR)\citep{huangfu_multi-task_2022} is brought in. Besides, Baltescu et al. \citet{baltescu_itemsage_2022} introduce a multimodal-based architecture to aggregate information from both text and image modalities, and utilize MTL to make ItemSage optimized for several engagement types.

\begin{table*}[htb]
\centering
\caption{A Summary of Multi-task Learning-based Models in E-commerce Recommendation}\label{tab1}
\begin{tabular}{p{0.15\linewidth}p{0.20\linewidth}p{0.12\linewidth}p{0.08\linewidth}p{0.08\linewidth}p{0.08\linewidth}p{0.08\linewidth}}
\hline
Recommender types & Models                                                    & Task1                & Task2                    & Task3 & Task4 & Task5 \\ \hline
E-commerce              & DUPN\citep{ni_perceive_2018}           & CTR                  & L2R            & PPP   & FIFP  & SPP   \\
E-commerce              & CrossDistil\citep{yang_cross-task_2022}  & Buy                  & Like             &       &       &       \\
E-commerce              & MLPR\citep{wu_multi-task_2022}        & CTR           & CVR             & ATC   &       &       \\
E-commerce              & M2TRec\citep{shalaby_m2trec_2022}     & Next-item  &  category &       &       &       \\
E-commerce              & TBCG\citep{luo_dual-task_2022}        & Next-item & Buy     &       &       &       \\
E-commerce              & MTDFM\citep{huangfu_multi-task_2022}   & CVR        & NDPR              &       &       &       \\
E-commerce              & \citep{baltescu_itemsage_2022}         & Buy          & ATC      & Save  & CTR &       \\
E-commerce              & HM\(^{3} \)\citep{wen_hierarchically_2021}      & CTR          & D-MI       & ATC  & CTCVR &       \\
E-commerce              & \citep{hadash_rank_2018}               & Ranking              & Rating                   &       &       &       \\ 
E-commerce             & ESMM\citep{ma_entire_2018}                 & CVR      & CTR         & CTCVR           & SPP      &    \\
E-commerce              & ESCM\(^{2} \)\citep{wang_escm2_2022}                & CVR      & CTR         & CTCVR           & SPP      &    \\
Video             & MFH\citep{liu_multi-faceted_2022}        & Cmpl     & Finish      & Skip            &       &       \\
E-commerce             &  NCS4CVR\citep{xiao_ncs4cvr_2023}& CVR      & CTR         &                 &       &       \\
E-commerce             &  MMOE\citep{ma_modeling_2018}& CVR      & CTR         &                 &       &       \\
E-commerce             &  PLE\citep{tang_progressive_2020}& CVR      & CTR         &                 &       &       \\
E-commerce             &  HEROES\citep{jin_multi-scale_2022}& CVR      & CTR         &                 &       &       \\
E-commerce             & \citep{zhang_large-scale_2020}                   & CVR      & CTR         &      &       &       \\
E-commerce             & ESM\(^{2} \)\citep{wen_entire_2020}          & CVR      & CTR         & CTAVR           & CTCVR &       \\ \hline
\end{tabular}
\end{table*}

For video recommendation, in addition to CTR, CVR and CTCVR, user feedback information such as Cmpl(Play Completion Ratio), finish(Play Finish Rate) and skip(Play Skip Rate) \citep{liu_multi-faceted_2022} can also be beneficial to recommendation. Especially, for content recommendation, Some natural language processing tasks will be added to assist the recommendation task, such as named entity recognition(NER) is often used in news recommendation.

\begin{table*}[htb]
\centering
\caption{A Summary of MTL-based Models in Advertising Recommendation}\label{tab2}
\footnotesize
\begin{tabular}{p{0.15\linewidth}p{0.12\linewidth}p{0.1\linewidth}p{0.12\linewidth}p{0.12\linewidth}p{0.12\linewidth}p{0.1\linewidth}}
\hline
Recommender types & Models                                             & Task1 & Task2    & Task3        & Task4      & Task5 \\ \hline
Advertising       &CTnoCVR\cite{zhang_ctnocvr_2022}       & CVR   & CTR      & CTnoCVR      &      &       \\
Advertising       & \cite{ma_online_2022}           & CTR   & Dismiss  & ClickQuality &            &       \\
Advertising       & MTAE\cite{yang_multi-task_2021} & CTR   & MP      & WP          &            &       \\
Advertising       & AITM\cite{xi_modeling_2021}     & CTR  & Buy  & Approval  & Activation &       \\
Advertising       & DINOP\cite{xin_multi-task_2019} & GMV   & BSP      & SS           & BOP        & JA    \\ \hline
\end{tabular}
\end{table*}

\subsection{Advertising Recommendation}

Advertising recommendation is a special type of E-commerce recommendation that aims to recommend ads that users may be interested in. So, CTR and CVR tasks are also used to recommend ads and CTR is the most critical task for advertising recommendation. For providing more accurate advertising recommendations, different from CTCVR task, Zhang et al. \citet{zhang_ctnocvr_2022} introduce a novelty auxiliary task CTnoCVR(the probability of action with click but no-conversion), which can promote samples with high CVR but low CTR. Moreover, user history behaviors and ad features are also essential for advertising recommendation, such as buy, approval, dismiss, ClickQuality(represents the probability that an ad click produce a good landing page experience)\citep{ma_online_2022}, Market Price(MP), Winning Probability(WP, represents the probability that the bid price is higher than the market price)\citep{yang_multi-task_2021}. 

Especially, aimed for online promotion recommendation, there are several particular tasks in \citep{xin_multi-task_2019}, such as Gross Merchandise Volume(GMV, indicates the total sales value), Best Selling Products (BSP, predicts whether one item in the list of best selling products), and Sale Slot (SS, means in-depth index of the BSP for each industry), Buying through Online Promotion(BOP, predicts whether the user will buy one item during the promotion) and Joining an Activity(JA, decides whether one item should be included in a promotional activity). Learning these tasks simultaneously can improve personalized recommendations in the online promotion. All tasks which are employed in advertising recommendation are listed in Table \ref{tab2}.

\begin{table}[btp]
\centering
\caption{A Summary of MTL-based Models in Explanation Recommendation}\label{tab3}
\scriptsize
\begin{tabular}{llll}
\hline
Recommender types & Models                                               & Task1                  & Task2               \\ \hline
Explanation       &\cite{lu_why_2018}             & Rating      & Review generation              \\
Explanation       & CAML\cite{chen_co-attentive_2019} & Rating &    Explanation generation           \\
Explanation       & ExpGCN\cite{wei_expgcn_2022}     & Item                   & Explanation ranking \\ 
Explanation       & J3R\cite{avinesh_j3r_2020}          &  Rating           &    Review summaries generation\\
Explanation       & ECR\cite{chen_towards_2020}         &    Rating        &   Explanation generation  \\\hline
\end{tabular}
\end{table}

\subsection{Explainable Recommendation}

Explainable recommendation is an emerging research direction that aims to provide users with the interpretability of the recommendation process so that they can better understand and accept the recommendation results. In Explainable recommendation, in addition to recommendation tasks, they also need introduce some explanation generation tasks to provide recommendation reasons to users, such as explanation generation and explanation ranking in Table \ref{tab3}. Lu et al. \citet{lu_why_2018} jointly learn rating prediction of a target user for an item and review generation which leverages adversarial sequence-to-Sequence learning to generate a review to serve as an explanation. Avinesh et al. \citet{avinesh_j3r_2020} provide a novel multi-task recommendation model which contains three components: user and item models, rating prediction, and review summary generation. Chen et al. \citet{chen_towards_2020} illustrate ECR model that learns cross knowledge for recommendation tasks and explanation simultaneously. 

\subsection{Conversational Recommendation}

Conversational recommendation refers to recommending appropriate responses or suggestions to users. However, there are little related work about multi-task conversational recommendation. UniMIND\cite{deng_unified_2022} provides a unified conversational recommender system which consists of four optimization goals, as shown in Table \ref{tab4}. For the sake of remember, we use abbreviations to represent these tasks, where GP means goal planning(selects the appropriate goal to determine where the conversation goes), TP means topic prediction(predicts the next conversational topics), IR means item recommendation(recommend an item based on the dialogue context and the user profile) and RG means response generation(generates a proper response for completing the selected goal).

\subsection{Multi-scenario Recommendation} \label{msr}

With the continuous evolution of recommender system, multi-scenario recommendation\cite{wang_causalint_2022} has emerged as a significant topic, leveraging information from different scenarios to enhance recommendation accuracy. Similar to parameter sharing of MTL, multi-scenario recommendation often employs shared modules and scenario-specific modules to extract both scenario-shared information and scenario-specific knowledge.

Multi-task and multi-scenario RSs \cite{yuan2024mmlrec} currently face three key challenges: 1. \textbf{Negative transfer}: The distribution and optimization goals of tasks across different scenarios are often inconsistent, or even conflicting, which can degrade the model's generalization performance. So, negative transfer is a key factor that needs to be carefully considered. 2. \textbf{Task/Scenario relationship modeling}: The relationships between different tasks and scenarios are complex, and capturing these inter-task associations to improve prediction performance across multiple scenarios remains a significant challenge. 3. \textbf{Cold start problem}: Insufficient data in new scenarios or tasks hinders the model's ability to provide accurate recommendations.

Regarding \textbf{challenge 1}, CSII \cite{shu2024adaptive} construct a transferable feature extraction module to identify highly transferable features among scenarios, which are subsequently provided to all scenario-specific experts. Chang et al. \citet{chang_pepnet_2023} aim to solve the imperfectly double seesaw phenomenon resulted by data distribution difference between different scenarios, and they take the personalized priors as the input of the gating network, then strengthen the personalization through the embedding personalized network module and parameter personalized network module.

Regarding \textbf{challenge 2}, Zou et al. \citet{zou_automatic_2022} integrate both MSL(multi-scenario learning) and MTL into a unified framework with an expert selection algorithm which can automatically select scenario-/task-specific and shared experts for each input. Li et al. \citet{li_improving_2020} propose multi-scenario learning to rank, which aims to deal with the difference of user behaviors among different scenario. They leverage MMoE model to identify differences and commonality between tasks in the feature space and learn scenario relationships in the label space with a stacked model. M3REC\cite{lan2023m3rec} leverages scenario correlations to overcome synchronous optimization issues and employs a task-specific aggregate gate for multiple recommendation tasks across scenarios. Some works take a more fine-grained approach to differentiate between scenarios/tasks at the feature level. MMFI \cite{song2024multi} introduces a novel scenario-task aware attention layer that learns the importance of feature interactions across different scenario and task pairs in a fine-grained manner. Additionally, it employs two sparsification operations to obtain sparse importance distributions, thereby eliminating redundant feature interactions.

Regarding \textbf{challenge 3},  Zhang et al. \citep{zhang_leaving_2022} exploit meta learning to capture inter-scenario correlations, which designs a meta attention module and a meta residual tower layer to learn the correlation between scenarios. The approach has been proven to scale easily to new scenarios. Zhou et al. \citet{zhou_hinet_2023} introduce hierarchical information extraction network to model task-related information and scenario-related information at different levels, which can capture complex relations of scenarios and tasks more efficient. Gong et al. \cite{gong2023unified} utilize LLM to extract domain invariant features, enabling effective feature sharing in cold-start scenarios. LLM4MSR\cite{wang2024llm4msr} first leverage LLM to uncover multi-level knowledge from the designed scenario-level and user-level prompt without fine-tuning the LLM, then adopt hierarchical meta networks to generate multi-level meta layers to explicitly improve the scenario-aware and personalized recommendation capability.

\subsection{Multi-domain Recommendation}\label{mdr}

Cross-domain recommendations \cite{man2017cross,li_recguru_2022} attract widespread attention, which have been proposed as a solution to the cold-start problem. This approach aims to transfer knowledge from a source domain with abundant data to a target domain with limited data, enabling effective recommendations despite data scarcity. In the past five years, methods that fuse MTL and cross-domain recommendation have gradually emerged. Zhu et al. \citet{zhu_dtcdr_2019} design an adaptable embedding sharing strategy to combine and share the embeddings of common users across domains based on the MTL. Bi et al. \citet{bi_mtrec_2022} propose a MTL framework to incorporate the multi-domain information into BERT, which improves its news encoding capability. Later, Gong et al. \cite{gong2023unified} propose a LLM-based multi-domain foundation model, incorporating a domain adaptive multi-task module to extract the domain-specific query and item towers’ representations. M\(^3\)oE\cite{zhang2024m3oe} is designed to address two key conflicts in multi-domain and multi-task recommendation scenarios: (1) the same cross-domain information transfer method may not generalize well across different tasks, and (2) the same multi-task optimization strategy may not adapt effectively to different domains. To tackle these challenges, M\(^3\)oE introduces a multi-view expert learning layer, which employs a disentangled design to capture domain-specific user preferences and task-specific user preferences effectively.  M3Rec\cite{lan2023m3rec} introduces a meta-item-embedding generator and a user-preference transformer to unify user and item representations across different domains. PRP \cite{wu2025prototype} utilizes a shared Mixture of Experts architecture to learn representations for each sample, projecting them into a common prototype space across all domains/tasks. HyperGate \cite{huang2025hypergate} augments semantic information for domains and tasks using contrastive learning based on MoE, which then serves as perceptive control for the subsequent hierarchical gating network.

\subsection{Other Recommendation}

In addition to above applications, researchers have started to utilize MTL for improving the performance of other recommendation tasks, such as social recommendation and travel recommendation. Since there are few studies about these recommendation applications, we will briefly introduce the related progress.

Social recommendation refers to recommending social relationships, social circles, social activities, etc. It is often used in social networks and community websites. SoNeuMF\citep{feng_social_2022} is proposed by sharing user representation in two tasks which are SR(social relation prediction, predict whether exist social relationship between two users) and interaction prediction(IP, predicts whether exist interaction between a user and an item). Besides, travel recommendation\citep{chen_multi-task_learning_2021} is also studied to provide travel recommendation(TR) and keywords generation(KG).

\subsection{Summary}

This section depicts six prominent applications of RSs, including E-commerce recommendation, advertising recommendation, explainable recommendation, conversational recommendation, multi-scenario recommendation, and cross-domain recommendation. Within these different recommendation application scenarios, there will be different MTL objectives, and existing work has made great progress in these fields. Notably, CTR and CVR have been the focus of numerous studies as they play a crucial role in influencing the efficacy of RSs, regardless of the application scenario.

Nonetheless, with the rising prominence of ChatGPT, conversational RSs gain increasing attention. This trend invites the consideration of whether a universal multi-task recommendation framework could be developed, one which would exhibit superior performance across an array of recommendation scenarios. Concurrently, the explainability of recommendations has garnered significant interest. Future research endeavors could place a stronger emphasis on harnessing the interrelationships among different tasks to unravel the underlying rationale of recommendations, thereby elucidating the operational tenets of certain opaque recommendation models.

\begin{table}[tbp]
\centering
\caption{A Summary of MTL-based Models in Other Recommendation}\label{tab4}
\scriptsize
\begin{tabular}{cccccc}
\hline
Recommender types & Models       & Task1   & Task2             & Task3 & Task4          \\ \hline
social           & SoNeuMF\cite{feng_social_2022}       & SRP       & IP   &    &      \\
conversational    & UniMIND\cite{deng_unified_2022} & GP         & TP        & IR  & RG    \\
travel       & TRKG\cite{chen_multi-task_learning_2021} &TR  & KG     &   &  \\ \hline
\end{tabular}
\end{table}

\section{Datasets and Evaluation}\label{sec8}

In this section, we introduce several commonly datasets and evaluation metrics for different recommendation tasks to help researchers find suitable datasets and evaluation metrics to test their MTL-based methods.

\subsection{Datasets}

One of the critical components in the development and evaluation of MTRSs is the availability of high-quality datasets. In recent years, to facilitate research in this area, various publicly available datasets have been developed for evaluating the performance of MTRSs models. This part introduces several public commonly datasets for MTRSs, as summarized in Table \ref{tab5}.

\begin{table*}[htp]
\centering
\caption{Statistics of multi-task benchmark datasets for recommendation tasks}\label{tab5}
\begin{tabular*}{\linewidth}{ccc}
\hline
Dataset            & Topic                                  & Website                                                                                            \\ \hline
Brightkite        &Social  & { \url{http://snap.stanford.edu/data/loc-brightkite.html}}                    \\ \hline
Ciao               & Social  & { \url{https://www.cse.msu.edu/$\sim$tangjili/trust.html}}                     \\ \hline
Ali-CCP            &       E-commerce                                 & {\url{https://tianchi.aliyun.com/datalab/dataSet.html?dataId=408}}                                         \\ \hline
Diginetica         &   E-commerce                  & {\url{https://competitions.codalab.org/competitions/11161}}                   \\ \hline
AliExpress2        & E-commerce                               & 
{\url{https://tianchi.aliyun.com/dataset/dataDetail?dataId=74690}}                                       \\ \hline
RetailRocket       &       E-commerce                                 & {\url{https://www.kaggle.com/datasets/retailrocket/ecommerce-dataset}}        \\ \hline
Tenrec\cite{yuan_tenrec_2022}       & Multi-scenario               & {\url{https://static.qblv.qq.com/qblv/h5/algo-frontend/tenrec\_dataset.html}} \\ \hline
Yelp2018           & Restaurant                          & { \url{https://www.yelp.com/dataset/challenge}}                                \\ \hline
Amazon-book        & Book                            & { \url{http://jmcauley.ucsd.edu/data/amazon}}                                  \\ \hline
Book-Crossing          &        Book                                & { \url{http://www2.informatik.uni-freiburg.de/~cziegler/BX/}}                                                         \\ \hline
Last.FM            & Music                                  & { \url{https://grouplens.org/datasets/hetrec-2011}}                            \\ \hline
TCM\citep{yao_topic_2018}                & Medicine                               &  { \url{https://github.com/yao8839836/PTM}}                             \\ \hline
TikTok             & Short-video                            & { \url{https://www.biendata.xyz/competition/icmechallenge2019/data/}}          \\ \hline
Kuairand           &         Short-video                               & { \url{https://kuairand.com/}}                                                                              \\ \hline
MovieLens          &     Movie                                  & { \url{https://grouplens.org/datasets/movielens/}}                                                          \\ \hline
Census-income Data &       Income                                 &           { \url{http://archive.ics.uci.edu/ml/datasets/Census-Income+\%28KDD\%29}}    \\ \hline
\end{tabular*}

\end{table*}

\subsection{Evaluation}

Selecting appropriate metrics is crucial for effectively evaluating the performance of MTRSs. Here we summarize some important evaluation metrics.

To facilitate a fair comparison of the performance of different methods, we have compiled the results of common MTRSs on CTR and CVR tasks (or CTCVR task) across five popular datasets, as presented in Table \ref{tab6}. This summary serves to provide a comprehensive overview of how these methods perform in real-world recommendation scenarios.

\begin{sidewaystable*}[p]
\caption{The performance comparison of different methods in common public datasets.}\label{tab6}

\begin{tabular}{ccccccccccccc}
\hline
\multirow{2}{*}{Public   Dataset} & \multirow{2}{*}{Task}  & \multirow{2}{*}{Metric} & \multicolumn{10}{c}{Methods}                                                                                                                                                  \\ \cline{4-13} 
                                  &                        &                         & ST & SB & MMoE            & PLE           & SS & CSRec            & ESMM             & AITM             & MNCM             & RMTL         \\ \hline
\multirow{4}{*}{Ali-CCP}          & \multirow{2}{*}{CTR}   & AUC \(\uparrow\)        & 0.6089      & 0.6098        & 0.6094          & \underline{0.6112}  & 0.6091         & 0.6098           & \textbf{0.6856}  & 0.6043           & -                & -                \\
                                  &                        & Gain\(\uparrow\)        & -           & +0.0009       & +0.0005         & \underline{ +0.0023} & +0.0002        & +0.0009          & \textbf{+0.0767} & -0.0046          & -                & -                \\
                                  & \multirow{2}{*}{CVR}   & AUC\(\uparrow\)         & 0.6011      & 0.6225        & 0.6223          & \underline{ 0.6355}  & 0.6063         & 0.6074           & 0.6338           & \textbf{0.6525}  & -                & -                \\
                                  &                        & Gain\(\uparrow\)        & -           & +0.0214       & +0.0212         & \underline {+0.0344} & +0.0052        & +0.0063          & +0.0327          & \textbf{+0.0514} & -                & -                \\ \hline
\multirow{4}{*}{AliExpress}       & \multirow{2}{*}{CTR}   & AUC\(\uparrow\)         & 0.7119      & 0.7123        & 0.7188          & \underline {0.7212}  & -              & -                & 0.7154           & 0.7128           & \textbf{0.7215}  & -                \\
                                  &                        & Gain\(\uparrow\)        & -           & +0.0004       & +0.0069         & \underline{ +0.0093} & -              & -                & +0.0035          & +0.0009          & \textbf{+0.0096} & -                \\
                                  & \multirow{2}{*}{CVR}   & AUC\(\uparrow\)         & 0.8647      & 0.8638        & 0.8677          & \underline {0.8692}  & -              & -                & 0.8663           & 0.8689           & \textbf{0.8716}  & -                \\
                                  &                        & Gain\(\uparrow\)        & -           & -0.0009       & +0.0030         & \underline {+0.0045} & -              & -                & +0.0016          & +0.0042          & \textbf{+0.0069} & -                \\ \hline
\multirow{4}{*}{CensusIncome}     & \multirow{2}{*}{CTR}   & AUC\(\uparrow\)         & 0.9433      & -             & 0.9477          & 0.9492        & \underline {0.9502}   & \textbf{0.9531}  & -                & -                & -                & -                \\
                                  &                        & Gain\(\uparrow\)        & -           & -             & +0.0044         & +0.0059       & \underline {+0.0069}  & \textbf{+0.0098} & -                & -                & -                & -                \\
                                  & \multirow{2}{*}{CVR}   & AUC\(\uparrow\)         & 0.9906      & -             & 0.9905          & 0.9917        & \underline {0.9921}   & \textbf{0.9933}  & -                & -                & -                & -                \\
                                  &                        & Gain\(\uparrow\)        & -           & -             & -0.0001         & +0.0011       & \underline {+0.0015}  & \textbf{+0.0027} & -                & -                & -                & -                \\ \hline
\multirow{8}{*}{RetailRocket}     & \multirow{4}{*}{CTR}   & AUC\(\uparrow\)         & 0.7273      & 0.7287        & \underline{ 0.7309}    & 0.7308        & -              & -                & 0.7282           & -                & -                & \textbf{0.7339}  \\
                                  &                        & Gain\(\uparrow\)        & -           & +0.0014       & \underline{ +0.0036}   & +0.0035       & -              & -                & +0.0009          & -                & -                & \textbf{+0.0066} \\
                                  &                        & Logloss\(\downarrow\)   & 0.2065      & 0.2048        & \underline{ 0.2021}    & 0.2056        & -              & -                & 0.2031           & -                & -                & \textbf{0.2013}  \\
                                  &                        & s-Logloss\(\downarrow\) & 0.0846      & 0.0839        & 0.0853          & \underline{ 0.0827}  & -              & -                & 0.0852           & -                & -                & \textbf{0.0824}  \\
                                  & \multirow{4}{*}{CTCVR} & AUC\(\uparrow\)         & 0.7250      & 0.7304        & 0.7347          & \underline{ 0.7387}  & -              & -                & 0.7316           & -                & -                & \textbf{0.7419}  \\
                                  &                        & Gain\(\uparrow\)        & -           & +0.0054       & +0.0097         & \underline{ +0.0137} & -              & -                & +0.0066          & -                & -                & \textbf{+0.0169} \\
                                  &                        & Logloss\(\downarrow\)   & 0.0489      & 0.0493        & 0.0496          & \underline{ 0.0486}  & -              & -                & \underline{ 0.0486}     & -                & -                & \textbf{0.0480}  \\
                                  &                        & s-Logloss\(\downarrow\) & 0.0150      & 0.0149        & \textbf{0.0145} & 0.0147        & -              & -                & 0.0150           & -                & -                & \underline{ 0.0146}     \\ \hline
\multirow{8}{*}{Kuairand}         & \multirow{4}{*}{CTR}   & AUC\(\uparrow\)         & 0.7003      & 0.7018        & 0.7014          & \underline{ 0.7026}  & -              & -                & 0.7009           & -                & -                & \textbf{0.7053}  \\
                                  &                        & Gain\(\uparrow\)        & -           & +0.0015       & +0.0011         & \underline{ +0.0023} & -              & -                & +0.0006          & -                & -                & \textbf{+0.0050} \\
                                  &                        & Logloss\(\downarrow\)   & 0.6127      & 0.6114        & 0.6119          & \underline{ 0.6111}  & -              & -                & 0.6128           & -                & -                & \textbf{0.6092}  \\
                                  &                        & s-Logloss\(\downarrow\) & 0.6263      & \underline{ 0.6250}  & 0.6255          & 0.6252        & -              & -                & 0.6261           & -                & -                & \textbf{0.6242}  \\
                                  & \multirow{4}{*}{CTCVR} & AUC\(\uparrow\)         & 0.7342      & 0.7310        & 0.7324          & 0.7339        & -              & -                & \underline{ 0.7350}     & -                & -                & \textbf{0.7367}  \\
                                  &                        & Gain\(\uparrow\)        & -           & -0.0032       & -0.0018         & -0.0003       & -              & -                & \underline{ +0.0008}    & -                & -                & \textbf{+0.0025} \\
                                  &                        & Logloss\(\downarrow\)   & 0.5233      & 0.5249        & 0.5237          & 0.5235        & -              & -                & \textbf{0.5220}  & -                & -                & \underline{ 0.5221}     \\
                                  &                        & s-Logloss\(\downarrow\) & 0.5449      & 0.5468        & 0.5450          & 0.5454        & -              & -                & \textbf{0.5436}  & -                & -                & \underline{ 0.5447}     \\ \hline 
\end{tabular}
\begin{tablenotes}
    \item Note: ST means single-task method, SB means shared bottom method, SS means sparse sharing method. \(\uparrow\) : higher is better; \(\downarrow\): lower is better. Bold: the best performance among all models. Underline: the second best model.
\end{tablenotes}
\end{sidewaystable*}

\textbf{Recall, F1, Precision} are widely used to evaluate the accuracy of recommendation results. Recall@K represents the proportion of recommended target item in the user's ground-truth set. Precision@K measures the proportion of target item in the top-K recommendation set. F1@K is the harmonic mean of Precision@K and Recall@K. The calculation formulas are as follows.
\begin{equation}
    \text{Recall@K}(u)=\dfrac{|R_u\cap G_u|}{|G_u|}
\end{equation}
where \(R_u\) is the set of recommended items predicted by the model, and \(G_u\) represents the ground-truth item set.
\begin{equation}
    \text{Precision@K}(u)=\dfrac{|R_u\cap G_u|}{K}
\end{equation}
\begin{equation}
    F1@K=\dfrac{2\times Precision@{}K\times Recall@{}K}{Precision@{}K+Recall@{}K}
\end{equation}
\textbf{AUC} is also a commonly used evaluation metric in RSs, which represents the probability that a clicked item is ranked in front of a non-clicked item. It is calculated as
\begin{equation}
    \operatorname{AUC}(u)=\dfrac{\sum_{i\in T(u)}\sum_{j\in I\setminus T(u)}I(\hat{r}_i>\hat{r}_j)}{|T(u)||L\setminus T(u)|}
\end{equation}
\textbf{Multi-AUC} (Multi-class Area Under ROC Curve) is adopt to measure multipartite ranking performance \citep{yang_cross-task_2022}, i.e.,
\begin{equation}
    \text { Multi-AUC }=\frac{2}{c(c-1)} \sum_{j=1}^{c} \sum_{k>j}^{c} p(j \cup k) \cdot AUC(k, j)
\end{equation}
where \textit{c} is the number of task classes, and \textit{p}(·) is prevalence-weighting function.

\textbf{GAUC} (Group AUC) is to calculate the AUC of each user, and then perform a weighted average. It is mostly used in advertising recommendations, which can better reflect the preference of different users for the ranking of different advertisements.
\begin{equation}
    GAUC=\frac{\sum_{(u,p)}w_{(u,p)}*A U C_{(u,p)}}{\sum_{(u,p)}w_{(u,p)}}
\end{equation}
\textbf{MSE} (Mean Squared Error) is a commonly adopt evaluation metric for regression tasks, which is used to measure the fitting degree of regression results. Besides, RMSE, MAE and R-Squared are also applied to evaluate the regression models.
\begin{equation}
    \operatorname{MSE}=\frac{\sum_{\operatorname{u.i}\in\operatorname{T}}(\operatorname{r_{ui}-\hat{r}_{\operatorname{ui}})^2}}{|\operatorname{T}|}
\end{equation}
\textbf{PR-AUC} (Precision-Recall Curve) is applied to balance precision and recall, which calculates the area under the Precision-Recall Curve.

\textbf{NDCG} (Normalized Discounted Cumulative Gain) evaluate the quality of the recommended top-K set generated by the model.
\begin{equation}
    \text{NDCG}@K=\dfrac{1}{|\mathcal{U}|}\sum_{u\in\mathcal{U}}\dfrac{\sum_{k=1}^K\dfrac{I(R_k^K(u)\in T(u))}{\log(k+1)}}{\sum_{k=1}^K\dfrac{1}{\log{(k+1)}}}
\end{equation}
\textbf{HR} (Hits Ratio) means the proportion of items that the user is interested in in the recommended top-K list.
\begin{equation}
    \text{HR@}K=\dfrac{1}{|\mathcal{U}|}\Sigma_{u\in\mathcal{U}}I(|R^K(u)\cap T(u)|>0)
\end{equation}
\textbf{MRR} (Mean Reciprocal Rank) represents the ranking of the target item in the actual recommended top-K list.
\begin{equation}
    MRR=\frac{1}{|U|}\sum_{i=1}^{|U|}\frac{1}{rank_i}
\end{equation}
\textbf{MTL gain} is proposed to measure the benefit of multi-task model compared with single-task models, it was proposed in \citep{tang_progressive_2020}, and its calculation formula is as follows.
\begin{equation}
    \text{MTL gain}=\begin{cases}M_{MTL}-M_{single},&M \text{ is a positive metric}\\ M_{single}-M_{MTL},&M \text{ is a negative metric}\end{cases}
\end{equation}
\textbf{MAP (Mean Average Precision)} is adopt as the measurement of prediction accuracy for each classification task and provide the average precision over users.
\begin{equation}
    \text{MAP}@K=\dfrac{1}{|\mathcal{U}|}\Sigma_{u\in\mathcal{U}}\Sigma^K_{k=1}\dfrac{I(R^K_k(u)\in T(u))\text{Precision}@k(u)}{K}
\end{equation}

Furthermore, we have summarize a comprehensive list of evaluation metrics adopted by different MTRSs, as shown in Table \ref{tab7}. The Analysis of Table \ref{tab7} reveals that among multi-task recommendation methodologies, the prevalent popular evaluation metrics encompass AUC, Recall/F1-score/Precision, and NDCG. The significance of these metrics within the multi-task recommendation domain reflects the common emphasis that different models place on prediction accuracy, retrieval quality, and recommendation relevance.
\begin{table*}[htb]
\centering
\caption{Evaluation metrics for multi-task recommendation tasks}\label{tab7}

\begin{tabular}{p{0.2\linewidth}p{0.8\linewidth}}
\hline
Metric & Models \\ \hline
AUC    & MMoE\citep{ma_modeling_2018}, ESMM\citep{ma_entire_2018}, HMoE\citep{li_improving_2020}, PLE\citep{tang_progressive_2020}, CrossDistil\citep{yang_cross-task_2022}, CSRec\citep{bai_contrastive_2022} , MLPR \citep{wu_multi-task_2022}, AESM\(^2\)\citep{zou_automatic_2022}, CFS-MTL\citep{chen_cfs-mtl_2022}, DFM-GCN\citep{xiao_dfm-gcn_2022}, DMTL\citep{zhao_distillation_2021},  HMKR\citep{wang_enhanced_2021}, ESCM2\citep{wang_escm2_2022}, FDN\citep{zhou_feature_2023}, GemNN\citep{fei_gemnn_2021}, HM\(^3\)\citep{wen_hierarchically_2021}, NCS4CVR\citep{xiao_ncs4cvr_2023}, RMTL\citep{liu_multi-task_2023}, MTAE\citep{yang_multi-task_2021}, HEROES\citep{jin_multi-scale_2022}, MFH\citep{liu_multi-faceted_2022}, AITM\citep{xi_modeling_2021}, \citep{xia_modeling_2018}, CTnoCVR\citep{zhang_ctnocvr_2022}\\ \hline
Multi-AUC    & CrossDistil\citep{yang_cross-task_2022}\\ \hline
MSE    &  CSRec\citep{bai_contrastive_2022}, CAML\citep{chen_co-attentive_2019}, FDN\citep{zhou_feature_2023}, PLE\citep{tang_progressive_2020}, NCS4CVR\citep{xiao_ncs4cvr_2023}, MFH\citep{liu_multi-faceted_2022}  \\  \hline
PR-AUC       &  MTDFM\citep{huangfu_multi-task_2022}, MoSE\citep{qin_multitask_2020}      \\  \hline
Recall/F1/Precision    &  GRU-MTL\citep{bansal_ask_2016}, DFM-GCN\citep{xiao_dfm-gcn_2022}, NextIP\citep{luo_dual-task_2022}, HMKR\citep{wang_enhanced_2021}, ESM\(^2\)\citep{wen_entire_2020}, ESCM2\citep{wang_escm2_2022}, ExpGCN\citep{wei_expgcn_2022}, FairSR\citep{li_fairsr_2022}, TMKG\citep{zhou_trust-aware_2022}, RnR\citep{hadash_rank_2018}, MTRec\citep{li_multi-task_2022}, M2GRL\citep{wang_m2grl_2020}, MedRec\citep{zhang_knowledge-enhanced_2022}\\  \hline
NDCG       &       MLPR\citep{wu_multi-task_2022}, NextIP\citep{luo_dual-task_2022}, ExpGCN\citep{wei_expgcn_2022}, MTER\citep{wang_explainable_2018}, FairSR\citep{li_fairsr_2022}, HMoE\citep{li_improving_2020}, TMKG\citep{zhou_trust-aware_2022}, NMTR\citep{gao_neural_2019}, MTRec\citep{li_multi-task_2022}, HEROES\citep{jin_multi-scale_2022}, MedRec\citep{zhang_knowledge-enhanced_2022} \\\hline
HR       &   TRKG\citep{chen_multi-task_learning_2021}, TMKG\citep{zhou_trust-aware_2022},  NMTR\citep{gao_neural_2019},  M2GRL\citep{wang_m2grl_2020}   \\ \hline
MRR       &   TRKG\citep{chen_multi-task_learning_2021}, RnR\citep{hadash_rank_2018}      \\ \hline
ROUGE       &    TRKG\citep{chen_multi-task_learning_2021}, CAML\citep{chen_co-attentive_2019}    \\ \hline
BLEU       &      CAML\citep{chen_co-attentive_2019}  \\ \hline
GAUC       &     HMoE\citep{li_improving_2020}, ESM\(^2\)\citep{wen_entire_2020},\\ \hline
MAP       &     SNR\citep{ma_snr_2019}, \citep{xia_modeling_2018}    \\ \hline
MTL gain       &     PLE\citep{tang_progressive_2020}, AITM\citep{xi_modeling_2021}    \\ \hline
Logloss      &    RMTL\citep{liu_multi-task_2023}, HEROES\citep{jin_multi-scale_2022}   \\\hline
\end{tabular}
\end{table*}

In addition, time complexity is also a important indicator to evaluate the efficiency and scalability of an algorithm. It provides insights into theoretical performance and helps compare the relative efficiency of different algorithms. In Table \ref{tab8}, we analyze the time complexity of several typical models, clarify their computational characteristics, and provide a comparative perspective for practitioners and researchers.
\begin{table}[h]
    \centering
\caption{The time and space complexity of some typical models}\label{tab8}
    \begin{tabular*}{\linewidth}{m{0.2\linewidth}m{0.3\linewidth}m{0.3\linewidth}} 
    \hline
         Model &Conference/Journal&  Time complexity\\ \hline
         MMoE &KDD'18&  \( O(dle+det) \)\\ 
         PLE &RecSys'20&  \( O(dlt) \)\\ 
          ESMM&SIGIR'18&  \(O(d+nl^2+l)\)\\ 
          ESM\(^2\)&SIGIR'20&  \(O(dl+nl^2+l)\)\\ 
          SNR&AAAI'19&  \( O(dst + n s^2 t) \)\\ 
          MSSM&SIGIR'21&  \( O(pdl+t(1-p)dl) \)\\
 MoSE& KDD'20& \( O(ds+tel^2s+tel) \)\\
 ESCM\(^2\)& SIGIR'22& \( O(d+nl^2+l) \)\\ 
\hline
    \end{tabular*}
    
   \begin{tablenotes}
    \item Note: \textit{d}: input dimension, \textit{e}: number of experts, \textit{t}: number of tasks, \textit{l}: hidden layer size, \textit{p}: Proportion of shared parameters(for MSSM model, a Multiple-level Sparse Sharing Model), \textit{s}: sequence length(for MoSE model, a Multitask Mixture of Sequential Experts model)
  \end{tablenotes}  
\end{table}

\section{Future Research Directions}\label{sec9}

Despite existing methods in MTRSs have received satisfactory results and established the foundation for MTRSs research, we think there remain several opportunities and challenges for future research. Rather than merely applying existing techniques to new domains, future research needs to address fundamental challenges related to the convergence of generative AI, the imperative for trustworthy systems, and the need for extreme scale and efficiency. In this section, we will discuss the future directions and challenges of MTRSs.

\subsection{Foundation Models and Generative MTRSs}

\paragraph{Why It's Important:}
The integration of foundation models, particularly large language models (LLMs), has become a transformative development in the field of MTRSs. LLMs bring immense potential to unify multiple recommendation tasks, such as rating prediction, item recommendation, explanation generation, and even conversational systems, into a single framework. This unification could alleviate the structural complexity of traditional multi-task recommendation models, which typically require task-specific architectures or “towers.” Furthermore, LLMs possess a vast amount of world knowledge and natural language understanding capabilities, which can address data sparsity and cold-start problems by leveraging pre-trained semantic knowledge across tasks. This makes LLM-based MTRSs especially crucial in real-world systems dealing with diverse and evolving data. Essentially, this represents a paradigm shift from "discriminative ranking" to "generative recommendation," promising a more universal and semantically rich approach to user modeling.

\paragraph{Challenges and Technical Difficulties:}
Despite the promising prospects, several challenges arise when integrating LLMs into MTRSs. One of the main difficulties is the heterogeneity of the tasks. Different tasks, such as click prediction and conversion prediction, have distinct objectives and data distributions, which can create task divergence when using a unified generative framework. Another challenge is maintaining the coherence and consistency of predictions across different tasks. While LLMs can handle multiple tasks, the generated output for one task may not align well with the needs of other tasks. A deeper technical challenge is the "semantic gap" between continuous, dense ID embeddings optimized for collaborative filtering and discrete semantic tokens used by LLMs. Aligning these two modalities without losing collaborative signals is non-trivial. Moreover, task coordination and data alignment in multi-task settings, especially with heterogeneous data types (e.g., user-item interaction vs. content data), require further refinement. Additionally, knowledge transfer between tasks needs to be more precisely calibrated, as improper transfer could lead to overfitting or bias in certain tasks. Finally, the inherent high latency of auto-regressive generation poses a severe bottleneck for real-time recommendation scenarios.

\paragraph{Future Research Directions and Current Exploration:}
Future work in this area should focus on the development of unified generative frameworks that can seamlessly combine multiple recommendation tasks into a single sequence-to-sequence generation problem. Techniques like prompt learning can be leveraged to effectively manage task coordination and semantic alignment. Furthermore, research should explore methods to fine-tune LLMs for specific domains while still leveraging their pre-trained knowledge to handle sparse and cold-start scenarios. To address computational costs, investigating Parameter-Efficient Fine-Tuning (PEFT) techniques like LoRA specifically adapted for multi-task recommendation is a critical direction. Additionally, the potential for multi-modal LLMs that can handle not just textual data but also image and video content, should be explored to address multi-modal recommendation tasks. Key existing works like P5 \citep{geng2022recommendation} and M6-Rec \citep{cui2022m6} provide foundational frameworks for these efforts. Future studies should investigate how to optimize these architectures for large-scale, multi-domain recommendation systems. Moreover, research in generative recommendation, as seen in works like MBGen \cite{tan2025implicit}, \cite{liu2024multi}, \cite{wang2025generative}, will be a new paradigm to modeling relationship f multiple behaviors in MTRSs.

\subsection{Trustworthy and Robust MTRSs}

\paragraph{Why It's Important:}
As MTRSs grow in complexity, ensuring their trustworthiness has become one of the most pressing issues. Trustworthiness in this context encompasses explainability, fairness, privacy protection, and robustness. MTRSs are inherently more complex than single-task systems, as they involve multiple objectives and tasks. This complexity raises the need for models that not only provide accurate recommendations but also offer transparent, interpretable, and fair outcomes. In a multi-task setting, a model must explain the trade-offs between conflicting goals (e.g., accuracy vs. diversity), ensure fairness across different tasks simultaneously, and protect privacy under complex shared parameter structures that introduce new attack surfaces. Additionally, privacy concerns are critical in today’s data-driven world, where sensitive information could be inferred from shared parameters in multi-task architectures. A trustworthy MTRS not only needs to perform well in terms of user satisfaction but also guarantee that the recommendations made are understandable, unbiased, and respect user privacy.

\paragraph{Challenges and Technical Difficulties:}
The primary challenge in building trustworthy MTRSs is managing task interference and conflicting objectives. For example, when balancing multiple tasks with different goals (such as maximizing clicks vs. minimizing clickbait), it becomes difficult to offer clear explanations of the model’s decision-making process. Ensuring fairness across tasks is another significant issue, especially when optimizing one task might introduce bias in others. For instance, enhancing fairness for one demographic group in the CTR task might worsen performance for another group in the CVR task. Achieving Pareto-optimal fairness across multiple conflicting objectives is a complex optimization challenge. Additionally, privacy preservation in MTRSs faces unique challenges due to shared representations, where one task may inadvertently leak sensitive data related to another via gradient inversion attacks targeting shared layers. Finally, robustness is another challenge, as MTRSs must not only perform well in typical scenarios but also be resilient to adversarial attacks, task perturbations, or shifts in user behavior.

\paragraph{Future Research Directions and Current Exploration:}
Research into conflict-aware explainability will be crucial for tackling the trade-offs made by MTRSs in balancing multiple objectives. Techniques such as causal inference and counterfactual reasoning \citep{li_relationship_2022} should be further developed to explicitly identify how and why certain compromises are made in the recommendations. On the fairness front, multi-objective Pareto optimization \citep{su2023task} will be a key research direction, as it provides a framework for optimizing fairness across all tasks simultaneously. Furthermore, leveraging adversarial learning to learn disentangled, task-invariant representations can inherently improve both fairness and robustness. Additionally, privacy-preserving MTL frameworks should be explored, incorporating techniques such as federated learning \citep{ammad_federated_2019, mei2024fedmoe} and differential privacy \citep{gao_dplcf_2020}, which prevent private data leakage while still allowing the model to learn shared representations. Research into adversarial robustness should also be a priority, ensuring that MTRSs can perform well even when faced with adversarially manipulated data. This will require the integration of robust learning techniques like adversarial training \citep{su2022cross} into multi-task recommendation models.

\subsection{Efficient and Scalable MTRSs}

\paragraph{Why It's Important:}
As MTRSs involve handling multiple tasks and datasets simultaneously, they inevitably face challenges in terms of scalability and computational efficiency. The ever-growing number of users and items, coupled with the increased complexity of multi-task models, requires highly efficient recommendation systems that can provide accurate results in real-time. The issue of parameter explosion exacerbates this challenge, as more tasks typically lead to more parameters, increasing both memory usage and inference latency. In high-concurrency industrial settings, achieving low latency while managing the enormous computational and energy costs of complex architectures like MoE is a fundamental barrier to deployment. In practical industrial settings, MTRSs must be scalable to handle massive datasets and able to generate real-time recommendations efficiently.

\paragraph{Challenges and Technical Difficulties:}
The primary difficulty in scaling MTRSs is the parameter explosion. As the number of tasks grows, the complexity of the model increases, resulting in an exponential rise in the number of parameters that need to be optimized. This makes real-time deployment infeasible without substantial computational resources. Specific to advanced architectures like MoE, the all-to-all communication overhead required to route samples to experts across distributed systems becomes a severe bottleneck. The second challenge is the dynamic nature of MTRSs, where tasks evolve and new tasks may be added frequently. Ensuring that the model adapts to these changes efficiently while maintaining high performance is another critical issue. Additionally, model distillation and network pruning techniques have been proposed as solutions to reduce model size and computational overhead, but their effectiveness in MTRSs still requires further investigation.

\paragraph{Future Research Directions and Current Exploration:}
Future research should focus on automated architecture search (NAS) for MTRSs to identify the best sharing topologies, minimizing the number of parameters while maximizing task performance. Dynamic inference techniques will also play a crucial role in improving efficiency by only activating the relevant sub-networks for each task. Sparse activation methods and early-exit mechanisms can further reduce computation and energy consumption, making MTRSs more efficient and suitable for real-time applications. This points towards "Green AI" approaches involving input-dependent conditional computation. Moreover, the integration of edge computing or distributed systems for MTRSs can help alleviate the computational burden of large-scale deployments. Finally, developing lifelong or continual MTL frameworks that allow models to learn new tasks incrementally without catastrophic forgetting is crucial for sustainable long-term deployment. Research into lightweight models for MTRSs is another promising direction, leveraging model distillation \citep{wang2023efficient} and parameter sharing across tasks to significantly reduce model size and inference time.

\section{Conclusion}\label{sec10}

With the increasing complexity of user behavior and diversification of decision-making processes, the traditional recommendation approaches reveal their limitations. MTL offer promise to address these limitations. Therefore, utilizing MTL techniques in RSs has obtained increasing interest in academia and industry. In this survey, we provide a systematically review of the state-of-art works on MTRSs. We analyzed from theoretical, empirical, and task relationship perspectives why MTL can deal with the core challenges of RSs. We organize existing works about MTRSs by model architecture, optimization strategy, training paradigm and application. Furthermore, we suggest some promising directions for the future research and we hope this survey can well assist students, staffs and experienced researchers in the relative areas.

To the researchers, practitioners, and enthusiasts involved in RSs, the current situation is full of challenges and opportunities. While significant strides have been made, there is scope for further exploration and innovation. In summary, the promise of MTL in improving and revolutionizing RSs is undeniable. Through this survey, we hope to have highlight its significance, illustrate its current state, and point the way to its future evolution.

\section*{Acknowledgments}

This work is partially supported by the R\&D Program of Beijing Municipal Education Commission (No. KM202310005028), and the Natural Science Foundation of Beijing,China (Grant No. 4244073).









\bibliographystyle{model1-num-names}


\bibliography{mtlrs}



\end{document}